\documentclass[journal]{IEEEtran}
\usepackage[utf8]{inputenc}
\usepackage{cite}
\usepackage{floatrow}
\usepackage{amsmath,amssymb,amsfonts,stfloats}
\usepackage{tabularx}
\usepackage{algorithmic}
\usepackage{graphicx}
\usepackage{textcomp}
\usepackage{floatrow}
\usepackage{xcolor}
\usepackage{booktabs}
\newtheorem{remark}{Remark}
\usepackage{multicol}
\usepackage{acronym}
\usepackage{url}
\usepackage{mathtools}
\usepackage[caption=false,font=footnotesize]{subfig}

\def\BibTeX{{\rm B\kern-.05em{\sc i\kern-.025em b}\kern-.08em
    T\kern-.1667em\lower.7ex\hbox{E}\kern-.125emX}}

\begin{document}
 
\title{ Optimized Waveform Design for OFDM-based ISAC Systems Under Limited Resource Occupancy }

\author{Silvia Mura,~\IEEEmembership{Member,~IEEE}, Dario~Tagliaferri,~\IEEEmembership{Member,~IEEE}, Marouan~Mizmizi,~\IEEEmembership{Member,~IEEE}, Umberto~Spagnolini,~\IEEEmembership{Senior Member,~IEEE}, and Athina~Petropulu,~\IEEEmembership{Fellow,~IEEE}
\thanks{This work was supported by the European Union under the Italian National Recovery and Resilience Plan (NRRP) of NextGenerationEU, partnership on “Telecommunications of the Future” (PE00000001 - program “RESTART”)}
\thanks{S.\, Mura. D.\, Tagliaferri, M.\, Mizmizi, U.\, Spagnolini are with the  Department of Electronics, Information and Bioengineering (DEIB) of Politecnico di Milano, 20133 Milan, Italy  (e-mail: [silvia.mura, dario.tagliaferri, marouan.mizmizi, umberto.spagnolini]@polimi.it }
\thanks{U.\, Spagnolini is Huawei Industry Chair}
\thanks{A.\, Petropulu is with Rutgers, the State University of New Jersey, NJ
08854, United States (e-mail: athinap@rutgers.edu).}}

\maketitle

\begin{abstract}

The sixth generation (6G) of wireless networks introduces integrated sensing and communication (ISAC), a technology in which communication and sensing functionalities are inextricably linked, sharing resources across time, frequency, space, and energy. 
Despite its popularity in communication, the orthogonal frequency-division multiplexing (OFDM) waveform, while advantageous for communication, has limitations in sensing performance within an ISAC network.
This paper delves into OFDM waveform design through optimal resource allocation over time, frequency, and energy, maximizing sensing performance while preserving communication quality.
During quasi-normal operation, the Base Station (BS) does not utilize all available time-frequency resources, resulting in high sidelobes in the OFDM waveform's ambiguity function as well as decreased sensing accuracy. To address these latter issues, the paper proposes a novel interpolation technique using matrix completion via Schatten $p$-quasi norm approximation, which requires fewer samples than the traditional nuclear norm for effective matrix completion and interpolation. This approach effectively suppresses sidelobes, enhancing sensing performance. Numerical simulations confirm that the proposed method outperforms state-of-the-art frameworks, such as standard complaint resource scheduling and interpolation, particularly in scenarios with limited resource occupancy.
\end{abstract}

\begin{IEEEkeywords}
Integrated sensing and communication, 6G, waveform design
\end{IEEEkeywords}

\maketitle

\section{Introduction}\label{sect:introduction}
6G is expected to be the first wireless generation to massively integrate radar sensing as a service thanks to new frequency bands (millimeter-wave (mmWave), $30-300$ GHz) and sub-THz ($> 100$ GHz), as well as the use of massive antenna arrays. Radar systems are widely utilized for various military and civilian applications, such as remote sensing, infrastructure monitoring, and driving assistance~\cite{8746875,PatoleAutomotiveRadars2020}, but they operate on dedicated spectrum portions to avoid interference. As 6G wireless networks demand large-scale and ubiquitous integration of radio sensing, equipping the communication infrastructure with standalone radars is not viable, as it would represent an unsustainable waste of hardware resources. In this regard, integrated sensing and communication (ISAC) systems emerged as a solution to the aforementioned problem, employing a single waveform for both communication and sensing functionalities over the same frequency/time/space and hardware resources~\cite{Heath2021overview,sun2020mimo}. Designing ISAC waveforms is challenging due to the different performance indicators for the two functionalities. Communication prioritizes reliable and high-capacity data transfer and orthogonal frequency division multiplexing (OFDM) waveforms to address frequency selective fading. On the other hand, radar systems focus on target detection and localization with sensing-optimal waveforms such as frequency-modulated continuous waveform (FMCW).  
To tackle the different requirements, a typical method for ISAC waveform is designed to optimize one functionality (either communication or sensing), while constraining the other to meet a certain quality of service (QoS),while there are also methods that consider waveforms that can trade-off the performance of one function for the other. This study considers a \textit{communication-centric} approach to ISAC design, where communication is the main goal and sensing is a secondary objective~\cite{Liu_survey}. In the following, we review the state of the art of waveform design for ISAC.

\subsection{Literature survey on ISAC waveform design}

Recent research in ISAC waveform design covers various domains such as space, frequency, and time. Information-theoretical approaches seek to bridge information and estimation theories. The ISAC waveform is designed to balance the maximization of channel capacity with the minimization of the Cramér-Rao bound (CRB) for estimation error of key sensing parameters\cite{10147248}. 
Useful inner and outer bounds on rate-CRB trade-off curve are reported.

From a more practical perspective, recent studies in \cite{liu2018toward,Sturm2011,Puccietal2022,Bica2019_multicarrier,xu2023bandwidth,9166743,Shi2018_powermin_OFDM_coexistence,9764299,zhang2023input,Yongjun2017_OFDMdesign_MI,Du2023,Wymeersch2021} focused on developing a single waveform that utilizes either space or time-frequency resources. In terms of spatial domain waveform design, initial ISAC efforts are focused on optimizing the beampattern across transmitting (Tx) antennas aiming at creating a beampattern suitable for both communication and sensing \cite{liu2018toward}.
Waveform design along time and frequency (and over the dual delay and Doppler (DD) domain) represented a major effort in the ISAC literature. 
Enhancing conventional OFDM waveform to improve sensing capabilities represents a cost-effective ISAC solution, ensuring retrofitting with 3GPP standards. Several studies, including~\cite{Sturm2011,Puccietal2022,Bica2019_multicarrier,xu2023bandwidth,9166743,Shi2018_powermin_OFDM_coexistence,9764299,zhang2023input,Yongjun2017_OFDMdesign_MI,Du2023,Wymeersch2021}, addressed OFDM-based ISAC design stemming from the standard-compliant 3GPP OFDM waveform. The seminal work in \cite{Sturm2011} was the first to suggest a signal processing algorithm for an OFDM-based radar. Work \cite{Puccietal2022} analyzes the ISAC performance capabilities of the 5G OFDM waveform, considering fully digital arrays and multi-beam design to split the spatial resources between communication and sensing. Work  \cite{Bica2019_multicarrier} suggests splitting OFDM subcarriers between radar and communication functionalities, where the performance trade-off between the two is implicit in the splitting ratio. Yet, incorporating subcarriers solely for radar function demands energy consumption that might be circumvented through a well-tailored allocation strategy catering to communication and sensing needs. Conversely, the OFDM waveform proposed in~\cite{xu2023bandwidth} balances communication efficiency and sensing performance by employing shared and private subcarriers. Maximizing the communication rate involves using all subcarriers as shared, whereas allocating more private subcarriers enhances sensing capabilities at the cost of the communication rate. The authors of \cite{9166743} propose super-resolution range and velocity estimators for OFDM-based ISAC systems.
The work in \cite{Shi2018_powermin_OFDM_coexistence} proposes three power minimization-based OFDM radar waveform designs for the coexistence between different radar and communication terminals on the same spectrum. A different power optimization method based on mutual information is explored in \cite{9764299}, which devises the power allocation strategy for communication-centric and radar-centric ISAC systems. The work in \cite{zhang2023input} considers the problem of reducing the probability of a wrong estimate of the range/velocity of a target due to a non-optimal input statistical distribution of communication symbols. The authors of \cite{Yongjun2017_OFDMdesign_MI,Du2023} employ information-theoretic metrics for communication and sensing channels to design the OFDM ISAC waveform. 
In \cite{Wymeersch2021} the authors consider the optimal resource allocation over time and frequency in an OFDM-based ISAC system, based on the proper minimization of delay and Doppler CRBs under communication constraints. Further constraints are set on the ambiguity function of the Tx signal, such that the sidelobes are kept within an acceptable level. 
All the previously mentioned studies focused on OFDM-based ISAC scenarios where all the time-frequency resources can be freely allocated. However, the waveform design approach (as well as the ISAC sensing algorithms) is markedly different in the case of underutilized resources (i.e., with a resource occupancy factor (ROF) $< 100\%$). In practical applications, the ROF is rarely near $100\%$, except in the case of severe traffic congestion in the network, as outlined in the 3GPP standard~\cite{rahman20215g} and corroborated by spectrum occupancy measurements campaigns~\cite{7460899}. 
Generally, the base station (BS) is designed to manage peak traffic, but for most of the time, it serves moderate traffic, which can result in low ROF levels. Low ROFs result in significant sidelobes within the ambiguity function, as depicted in Fig.\ref{fig:scenario}, that detrimentally affect sensing capabilities, calling for proper countermeasures~\cite{Wymeersch2021}. Works attempting to achieve low sidelobes of the ambiguity function with limited resources are in \cite{BarnetoFullDuplex,Barneto2021_Optimized_Wave}. The work~\cite{BarnetoFullDuplex} addresses practical considerations and challenges regarding delay/Doppler estimation using the 5G OFDM waveform with unused resources and it introduces a linear interpolation technique to reconstruct the sensing channel, from which to estimate the delay and Doppler of targets. A leap forward has been made in~\cite{Barneto2021_Optimized_Wave}, where the authors propose to fill the empty communication subcarriers with sensing pilots (i.e., radar subcarriers). The power and phase of radar subcarriers are optimized by minimizing the CRB on the delay and Doppler estimation for a single target, while limiting the peak-to-average power ratio. Our previous work~\cite{10264814} proposes to superpose to the standard-compliant time-frequency OFDM signal (with a variable number of occupied resources) a purposely designed low-power sensing signal with the desired ambiguity function. 
More recently, orthogonal time-frequency-space (OTFS) modulation has been investigated for ISAC purposes. Unlike OFDM, OTFS places data symbols in the DD domain, addressing issues in doubly-selective channels~\cite{Saif2021_OTFS,li2023isac,yuan2022orthogonal}. However, integrating OTFS into current 3GPP standards requires substantial modification due to its burst processing of consecutive OFDM symbols, which conflicts with the low latency requirements of many 6G services \cite{Saif2021_OTFS}.

\subsection{Contributions}

In light of the aforementioned literature, this work focuses on ISAC waveform design over time, frequency, and energy under limited resource occupancy constraints. We substantially extend our previous work~\cite{mura2023waveform}, where the waveform design is considered only across time and frequency. Herein, we introduce a further degree of freedom, (the energy allocation), and we also detail the sufficient conditions for a reliable sensing channel interpolation technique, guided by the isometric condition and the relative well-conditionedness property. Furthermore, the impact of the chosen p-value for the Schatten-$p$ quasi-norm interpolation is examined.
The main contributions can be summarized as follows:

\begin{itemize}
    \item We present a novel ISAC waveform design for OFDM-based systems that, unlike state-of-the-art methods, considers limited frequency-time resource occupancy, closely mimicking the resource usage of realistic applications. %This waveform serves both sensing and communication purposes using the same resources. 
    The design is achieved through an optimization problem that minimizes the weighted sum of CRBs for delay and Doppler estimation in the general case of two coupled targets, while adhering to achievable rate and time-frequency resource occupation constraints. Two waveform design method are proposed: \textit{(i)} waveform design by time and frequency resource allocation for fixed energy spectral density, and \textit{(ii)} waveform design over time, frequency, and energy. The advantages and disadvantages of both approaches are thoroughly discussed.
    \item In contrast to the state-of-the-art methods, which constrain the sidelobe level within the waveform optimization process, we address the sub-optimal ambiguity function issue arising from low ROF by establishing a framework for delay-Doppler parameter estimation based on sensing channel interpolation. Initially, the ISAC waveform is designed according to CRB minimization. Then, the sensing channel is estimated using a maximum likelihood approach over the available time-frequency resources. Finally, the sensing channel is interpolated via Schatten $p$-quasi norm matrix completion to minimize the sensing channel rank. The choice of the Schatten $p$-quasi norm is specifically targetet to the considered interpolation problem, as it requires fewer samples compared to the traditional nuclear norm. The conditions for the unique recovery of the sensing channel are aso provided.

    \item A comparative analysis evaluates the proposed ISAC waveform performance against conventional OFDM ones, including standard-compliant random resource scheduling and contiguous scheduling. Benchmarks operate with a fixed ROF and constant energy per resource and employ linear interpolation as outlined in~\cite{BarnetoFullDuplex}. The proposed ISAC waveform significantly outperforms the benchmarks, demonstrating 6× and 14× CRB gains under limited ROF. Moreover, under these conditions, the proposed approach successfully achieves CRBs for delay and Doppler estimation at high signal-to-noise ratios (SNR). In contrast, the method described in \cite{BarnetoFullDuplex} fails due to under-sampling. 

\end{itemize}

\textit{Organization}: 
The paper is structured as follows: Section \ref{sect:system_model} introduces the system model, while Section \ref{sect:metrics} explains the communication and sensing performance metrics for waveform design. Section \ref{sect:TF_scheduling} focuses on time and frequency waveform design, and Section \ref{sect:TFP_scheduling} outlines the general time-frequency-energy method. Sensing channel interpolation and parameter estimation are discussed in Section \ref{sect:interp}, and simulation results are presented in Section \ref{sect:numerical_results}. Finally, Section \ref{sect:conclusion} concludes the paper.

\textit{Notation}: 
The paper employs the following notation: Bold uppercase and lowercase letters represent matrices and column vectors, respectively. The $ij$-th entry of matrix $\mathbf{A}$ is denoted as $[\mathbf{A}]_{ij}$. Transposition, conjugate transposition, and $L$-quasi norm of matrices are represented by $\mathbf{A}^T$, $\mathbf{A}^H$, and $|\mathbf{A}|_L$, respectively. The element-wise product of matrices is denoted by $\odot$. $\mathrm{diag}(\mathbf{A})$ extracts the diagonal of matrix $\mathbf{A}$, while $\mathrm{vec}(\mathbf{A})$ represents vectorization by columns and $\mathrm{vec}^{-1}(\cdot)$ denotes the inverse operation. $\mathbf{1}_N$ is a column vector with $N$ entries equal to one. $\mathbf{a}\sim\mathcal{CN}(\boldsymbol{\mu},\mathbf{C})$ denotes a circularly complex Gaussian random variable with mean $\boldsymbol{\mu}$ and covariance $\mathbf{C}$. $\mathbb{E}[\cdot]$ is the expectation operator, and $\mathbb{R}$, $\mathbb{C}$, and $\mathbb{B}$ denote the sets of real, complex, and Boolean numbers, respectively. $\delta_{n}$ represents the Kronecker delta, where $\delta_{n-n'} = 1$ only if $n = n'$.

\section{System Model}\label{sect:system_model}

\begin{figure}[t!]
    \centering
    \includegraphics[width=0.9\columnwidth]{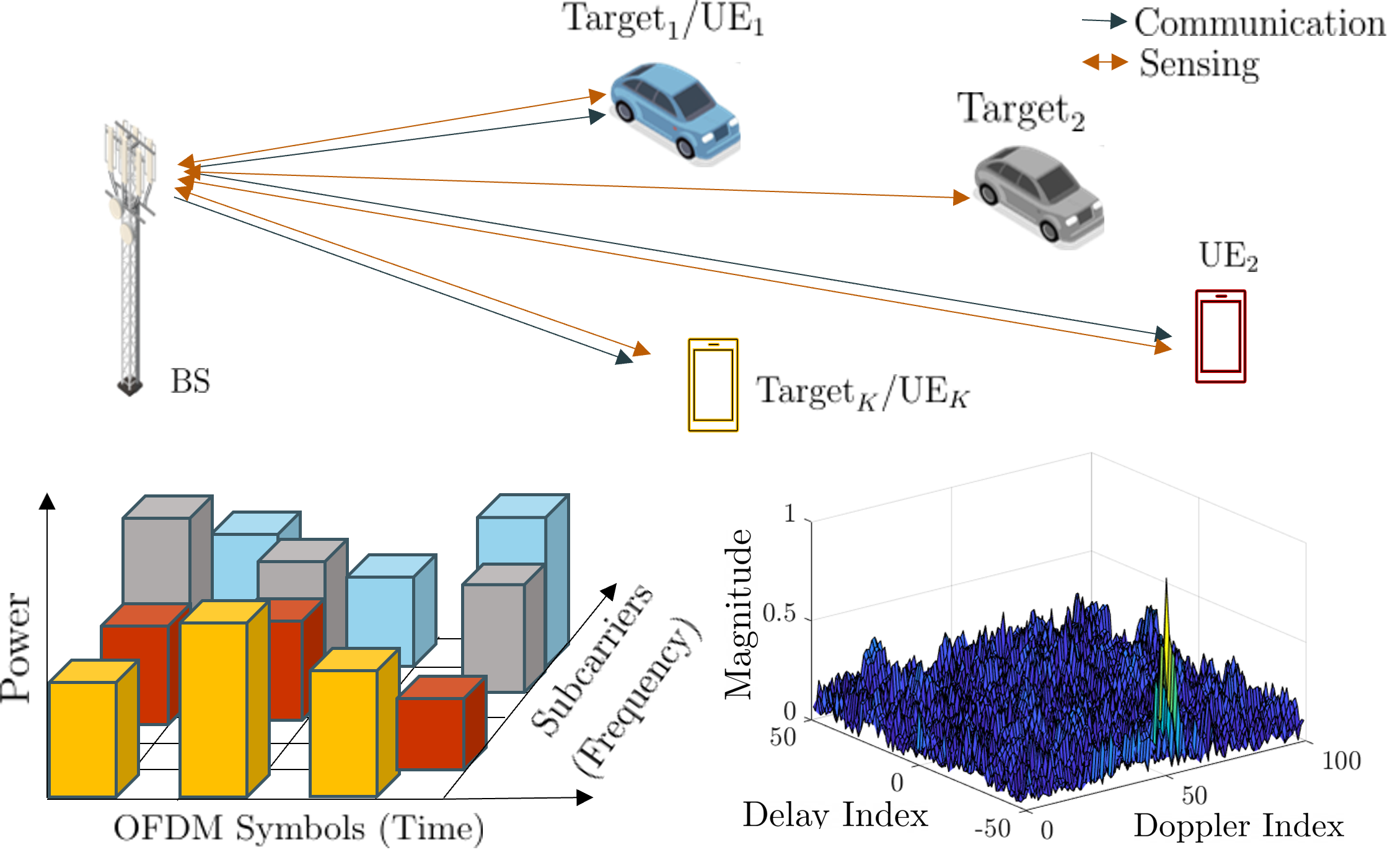}
     \caption{The ISAC waveform and its corresponding 2D ambiguity function are plotted within a scenario involving multiple targets/UEs. Unused resources result in high sidelobes in the ambiguity function, potentially deteriorating the estimation performance of sensing targets.}\label{fig:scenario}
\end{figure}

We consider the ISAC system illustrated in Fig. \ref{fig:scenario}, where the primary task of the multiantenna BS is to serve $K$ single antenna users' equipment (UEs) while simultaneously sensing the environment. For the sake of simplicity, the UEs are considered as the exclusive targets within the scenario. However, it's worth noting that the design principles herein can be extended to scenarios comprising multiple targets, whether separate or conjoined with the UEs. This work centers on the design of the ISAC waveform in the time-frequency domain, thereby implying that any spatial precoding/decoding at the transmitting/receiving antennas can be easily included in the approach. The ISAC BS utilizes an OFDM waveform where the time and frequency resources designated for downlink communication and sensing tasks are partitioned into $M$ subcarriers, spaced by $\Delta f$, and $N$ time slots (OFDM symbol duration $T = 1/\Delta f$). The overall bandwidth is $B = M \Delta f$, while the duration of the ISAC burst is $NT$. 
The Tx signal over time and frequency is represented by a matrix $\mathbf{X} \in \mathbb{C}^{M \times N}$, whose $mn$-th element is 
\begin{equation}\label{eq:tx_power}
    [\mathbf{X}]_{mn} = [\boldsymbol{\Sigma} \odot \mathbf{S}]_{mn} = \sigma_{mn} s_{mn},
\end{equation}
where $[\boldsymbol{\Sigma}]_{mn}= \sigma_{mn}\in \mathbb{R}_+$ is the square root of the allocated energy to the $mn$th communication symbol, denoted as $[\mathbf{S}]_{mn}=s_{mn}\in \mathbb{C}$. The symbol $s_{mn}$ is drawn from an arbitrary constellation with unitary energy such that $\mathbb{E}[s_{mn}]=0, \mathbb{E}[|s_{mn} s_{m'n'}|^2]= \delta_{m-m'}\delta_{n-n'}$, thus the total energy of the Tx signal is $E = \| \mathbf{X}\|_F^2 = \| \boldsymbol{\Sigma}\|_F^2$.

\subsection{Received signal at the ISAC BS}

The received sensing signal matrix $\mathbf{R} \in \mathbb{C}^{M \times N}$ at the BS in the time-frequency domain is 
\begin{equation}\label{eq:RxsignalBS}
    \mathbf{R} = \mathbf{X} \odot \mathbf{H}_s + \mathbf{W},
\end{equation}
where $\mathbf{H}_s \in \mathbb{C}^{M \times N}$ represents the sensing channel capturing the echos from the $K$ UEs/targets and $\mathbf{W} \in \mathbb{C}^{M \times N}$ gathers the noise samples within the frequency-time domain, such that $[\mathbf{W}]_{mn}=w_{mn} \sim\mathcal{CN}(0, N_0 \delta_{m-m'}\delta_{n-n'})$ and it is statistically uncorrelated across time and frequency. The sensing channel $\mathbf{H}_s \in \mathbb{C}^{M \times N}$ pertaining to the $mn$th resource bin is 
\begin{align}\label{eq:sensing_channel_FT}
 [\mathbf{H}_s]_{mn} = \sum_{k=1}^{K} \beta_{k} \, e^{j 2 \pi (\nu_{k} n T-\tau_{k} m \Delta f)},   
\end{align} 
where $(i)$ $\beta_{k} \sim \mathcal{CN}(0,\Omega^{(k)}_{\beta})$ denotes the complex scattering amplitude associated with the $k$-th UE/target with $\Omega^{(k)}_{\beta}$ proportional to $f_0^{-2} R_k^{-4}$ and contingent upon the carrier frequency $f_0$, the distance $R_k$ between BS and $k$th UE/target, and the reflectivity of the target, $(ii)$ $\tau_{k}=2 R_k/c$ corresponds to the propagation delay related to the $k$-th UE/target, $(iii)$ $\nu_{k}= 2 f_0 V_k/c$ represents the Doppler shift arising from the radial velocity $V_k$ associated with the $k$-th UE/target. Equations \eqref{eq:RxsignalBS} and \eqref{eq:sensing_channel_FT} hold valid under the assumption that the maximum delay, denoted as $\tau_{max} = \mathrm{max}_k\left(\tau_k\right)$, is constrained to be less than the duration of the employed cyclic prefix $T_{cp}$, ensuring $\tau_{max}\leq T_{cp}$ to enable unambiguous range estimation. 

\subsection{Received signal at the UE}

The time-frequency received signal at the $k$-th UE within the $mn$-th frequency-time resource bin is 
\begin{align}
      \mathbf{Y}_k = \mathbf{X} \odot \mathbf{H}_{k} + \mathbf{Z},  
\end{align}
and the communication channel is defined as
\begin{align}
 [\mathbf{H}_k]_{mn} =\sum_{q=1}^{Q} \alpha_{q}^{(k)} \,e^{j 2 \pi( {\nu}_{q}^{(k)} nT - m\Delta f {\tau}_{q}^{(k)})},   
\end{align} 
where $Q$ denotes the number of paths, considered uniform across all UEs for the sake of simplicity. Within each $q$-th path, $\alpha_{q}^{(k)} \sim \mathcal{CN}(0,\Omega^{(k)}_{q})$ represents the complex amplitude associated with the $k$-th UE. The parameters ${\tau}_{q}^{(k)}$ and ${\nu}_{q}^{(k)}$ signify the delay and Doppler shift, respectively, of the $q$-th path of the $k$-th UE. Unlike the sensing receiving signal $\mathbf{R}$ at the BS, the communication channel cannot retain the authentic delay and Doppler shifts due to time-frequency synchronization carried out by the UE terminal. The additive noise denoted as $z^{(k)}_{mn} \sim\mathcal{CN}(0, N_0 \delta_{m-m'}\delta_{n-n'}\delta_{k-\ell})$, remains uncorrelated across UEs as well as across time and frequency.

\section{Design Metrics for ISAC Waveform}\label{sect:metrics}

The paper aims to develop a dual-functional time-frequency waveform optimizing both sensing and communication capabilities. This section introduces metrics to evaluate the proposed waveform's performance in the ISAC context, categorized into sensing and communication metrics.

\subsection{Sensing Metrics}

The sensing performance is quantified by defining the CRB for the estimated delay and Doppler shift of the $K$ UEs/targets. The derivation adheres to \cite{Wymeersch2021}. We consider, for simplicity, the estimation of delay and Doppler shifts only, while the scattering amplitudes $\{\beta_k\}_{k=1}^K$ are known. This assumption does not limit the technical extent of the work (as the scattering amplitudes can be included in the CRB derivation) but it eases the analytical derivations. The CRB evaluation proceeds by vectorizing \eqref{eq:RxsignalBS} as follows:
\begin{equation}\label{eq:vec_rxsignal}
\mathbf{r} = \mathrm{vec}(\mathbf{R}) = \mathbf{x} \odot \sum_k \beta_k \odot (\mathbf{d}_{\tau,k} \otimes \mathbf{d}_{\nu,k}) + \mathbf{w},
\end{equation}
where 
\begin{align}\label{eq:tau_nu_vec}
    \mathbf{d}_{\tau,k} & = \left[e^{j\pi M \Delta f \tau_k} ,...,1,...,e^{-j\pi (M-1)\Delta f \tau_k}\right]^T\\
    \mathbf{d}_{\nu,k} & = \left[e^{-j\pi N T \nu_{k}} ,...,1,...,e^{j\pi (N-1) T \nu_{k}}\right]^T.
\end{align}
denote the delay and Doppler sensing channel responses of the $k$th target, while $\boldsymbol{\theta} = [\tau_1,...,\tau_K,\nu_1,....,\nu_K]^T\in\mathbb{R}^{2K\times 1}$ denotes the vector of parameters to be estimated. The Fisher information matrix (FIM) is block partitioned:
\begin{equation}\label{eq:fisher}
 \mathbf{F} = \begin{bmatrix}
     \mathbf{F}_{\mathbf{\tau}} \,\mathbf{F}_{\mathbf{\tau}\mathbf{\nu}}\\ \mathbf{F}^{T}_{\mathbf{\tau}\mathbf{\nu}}\,\mathbf{F}_{\mathbf{\nu}} 
 \end{bmatrix}
\end{equation}
and the $k\ell$th entry of each of the partitions is reported in  \eqref{eq:CRB_tau1tau2}-\eqref{eq:CRB_tau1nu2}, where $\mathbf{e} = \sum_k \mathbf{e}_k \in \mathbb{R}_{+}^{NM \times 1}$ is the vector of overall energy allocated to every single resource, while $\mathbf{e}_k$ refers to the vector of allocated energy per resource for the $k$th UE. The overall energy is $E=\mathbf{1}^T \mathbf{e}$.
\begin{figure*}
\begin{equation}\label{eq:CRB_tau1tau2}
    [\mathbf{F}_{\tau}]_{k,\ell} = \frac{2}{{N_0}}\, \mathbf{e}^T\Re\left\{4\pi^2 \beta_k\beta_\ell^*\Delta f^2 (\mathbf{m}\hspace{-0.05cm}\odot\hspace{-0.05cm}\mathbf{m}\odot \mathbf{d}_{\tau,k \ell} ) \otimes \mathbf{d}_{\nu,k \ell}\right\} = \mathbf{e}^T \boldsymbol{\zeta}_{\tau}^{(k,\ell)}
\end{equation}
\begin{equation}\label{eq:CRB_nu1nu2}
    [\mathbf{F}_{\nu}]_{k,\ell}= \frac{2}{N_0} \,\mathbf{e}^T \Re\left\{\mathbf{d}_{\tau,k \ell} \otimes 4\pi^2 T^2 \beta_k\beta_\ell^*(\mathbf{n}\odot\mathbf{n})\odot \mathbf{d}_{\nu,k \ell} \right\} = \mathbf{e}^T \boldsymbol{\zeta}_{\nu}^{(k,\ell)}
\end{equation}
\begin{equation}\label{eq:CRB_tau1nu2}
    [\mathbf{F}_{\nu,\tau}]_{k,\ell} = \frac{2}{N_0} \mathbf{e}^T \Re\left\{
	 4 \pi^2 \beta_k \beta^*_\ell \Delta f \left(\mathbf{m}\odot \mathbf{d}_{\tau,k \ell} \right) \otimes T \left(\mathbf{n}\odot \mathbf{d}_{\nu,k \ell}\right)    
    \right\} = \mathbf{e}^T \boldsymbol{\zeta}_{\tau,\nu}^{(k,\ell)}
\end{equation}
\hrulefill
\end{figure*}
In \eqref{eq:CRB_tau1tau2}-\eqref{eq:CRB_tau1nu2}, 
\begin{align}
    \mathbf{n}=\left[-\frac{N}{2},...,\frac{N}{2}-1\right]^T,\,\,\,
    \mathbf{m}=\left[-\frac{M}{2},...,\frac{M}{2}-1\right]^T
\end{align}
are the time and frequency index of the resource grid, while
\begin{equation}\label{eq:}
\begin{split}
    \mathbf{d}_{\tau,k\ell} = \mathrm{diag}(\mathbf{d}_{\tau,k}\mathbf{d}_{\tau,\ell}^H),\,\,\,\,\,
    \mathbf{d}_{\nu,k\ell} = \mathrm{diag}(\mathbf{d}_{\nu,k}\mathbf{d}_{\nu,\ell}^H)
\end{split}
\end{equation}
are the cross-coupled delay and Doppler channel responses, depending, respectively, on the delay difference $\tau_k-\tau_\ell$ and on the Doppler difference $\nu_k-\nu_\ell$. 
Under the assumption of $\mathbf{F}_{\nu}$ and $\mathbf{F}_{\tau}$ being non-singular, the CRB for delay and Doppler estimation is:
\begin{align}
    \mathbf{C}_\tau & = (\mathbf{F}_{\tau}-\mathbf{F}_{\nu,\tau}^{T}\mathbf{F}_{\nu}^{-1} \mathbf{F}_{\nu,\tau})^{-1}\label{eq:CRB_delay}\\
    \mathbf{C}_\nu & = (\mathbf{F}_{\nu}-\mathbf{F}_{\nu,\tau}^{T}\mathbf{F}_{\tau}^{-1} \mathbf{F}_{\nu,\tau})^{-1}.\label{eq:CRB_Doppler}
\end{align}

In practical scenarios, delay and Doppler estimation of targets are considered decoupled if their differences exceed system resolution ($|\tau_2-\tau_1| \gg \Delta \tau =1/B$ and $|\nu_2-\nu_1| \gg \Delta \nu =1/(NT)$). In such cases, the estimation of one target is unaffected by others, and CRB for multiple targets reduces to that of a single target, achieving minimum value. However, targets are often coupled due to dense environments or limited resources as the ISAC system usually observes extended targets. While the waveform design problem considers a generic number of targets $K$, the focus here is on two coupled targets ($K=2$) in numerical results.  Closed-form expressions for delay and Doppler CRBs for $K=2$ coupled targets are provided in Appendix \ref{app:CRB}.

\subsection{Communication Metrics}

The communication performance is quantified in terms of the achievable rate over each frequency-time resource. The SNR pertaining to the $k$th UE, $\ell$th resource is
\begin{equation}\label{eq:SNR_true}
    \gamma_{k,\ell} = \frac{[\mathbf{e}]_{\ell} \, \lvert [\mathbf{h}_k]_\ell \rvert^2 }{N_0 },
\end{equation}
where $\mathbf{h}_k = \mathrm{vec}(\mathbf{H}_k)$ is the $k$th communication channel vector. We consider the average achievable rate as the communication metric for the waveform design method, yielding:
\begin{align}
\eta_k = \frac{1}{L}\sum_{\ell=1}^L\log_2(1+\gamma_{k,\ell}).
\end{align}
Notice that  \eqref{eq:SNR_true} assumes perfect channel knowledge at the ISAC BS side. This information usually comes from channel state information reporting from the UEs and affects the waveform design method, as detailed in Section \ref{sect:TF_scheduling}. However, since the goal of this paper is to present a waveform design possibly independent of the individual realization of the communication channel, herein, we assume that the ISAC BS only knows the average channel gain over the resources, namely $ |H_{k}|^2 = \mathbb{E}_\alpha \left[\| \mathbf{H}_k \|_F^2\right]/L$. In this way, the SNR per resource becomes
\begin{equation}\label{eq:SNR_approx}
    \gamma_{k,\ell} = \frac{[\mathbf{e}_k]_{\ell} \, |H_{k}|^2 }{N_0 }
\end{equation}
and the time-frequency waveform is optimized on the mean and does not depend on the instantaneous channel realization. Of course, the proposed waveform design methods apply to \textit{any} communication channel, by using \eqref{eq:SNR_true} for the individual realizations.

\section{Time-Frequency Optimization}\label{sect:TF_scheduling}

\begin{figure}[t!]
\centering
 \subfloat[$\mu  = 0.25$]{ \includegraphics[width=0.5\columnwidth]{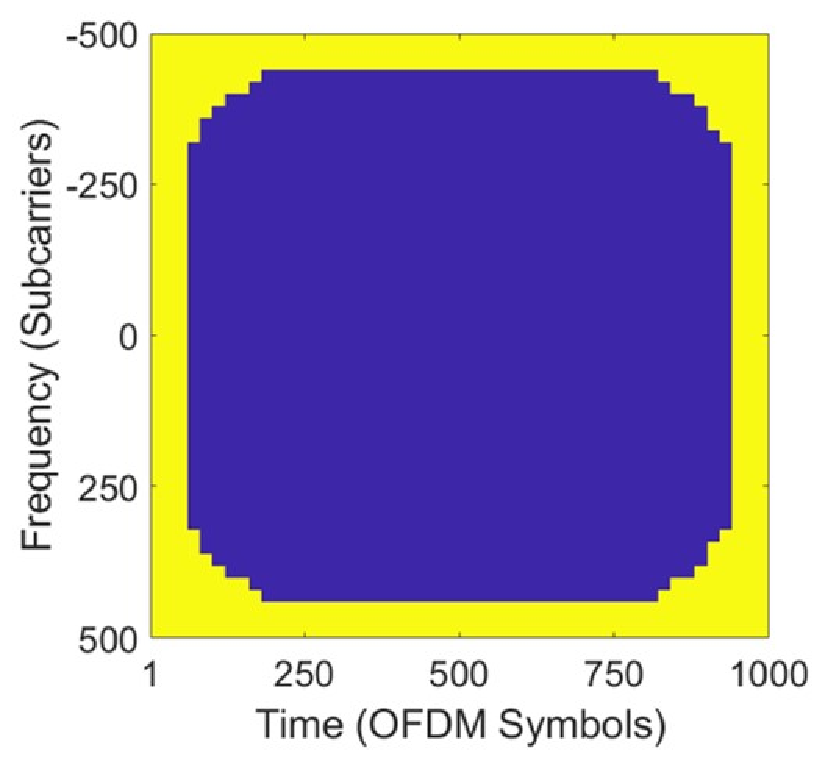}\label{fig:waveformb1}}
 \subfloat[$\mu  = 0.50$]{\includegraphics[width=0.5\columnwidth]{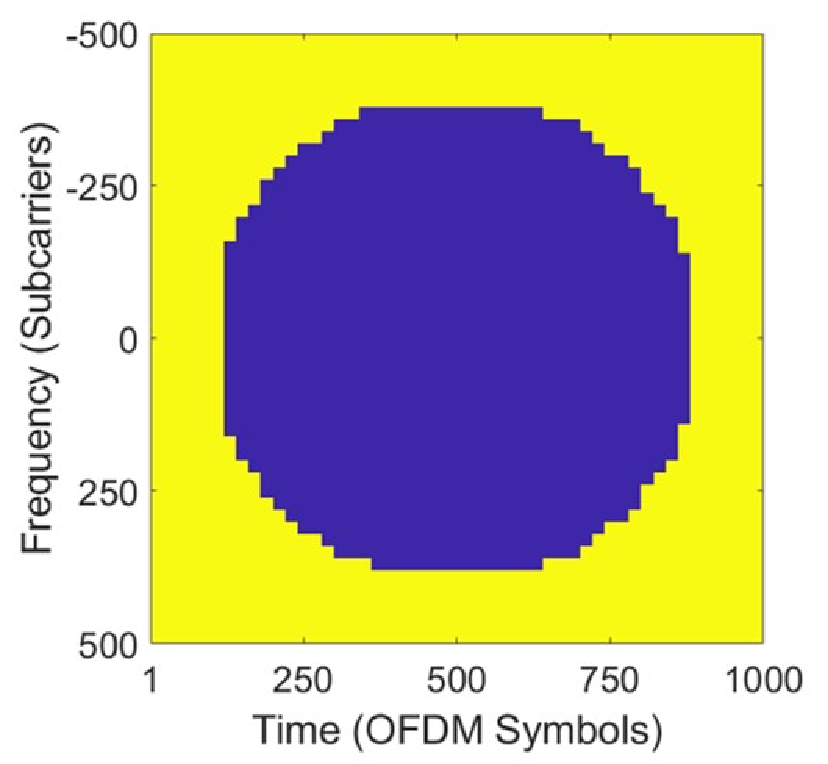}\label{fig:waveformb2}}\\
  \subfloat[${\epsilon}_{\tau} = 0.25,{\epsilon}_{\nu}= 0.75$]{ \includegraphics[width=0.5\columnwidth]{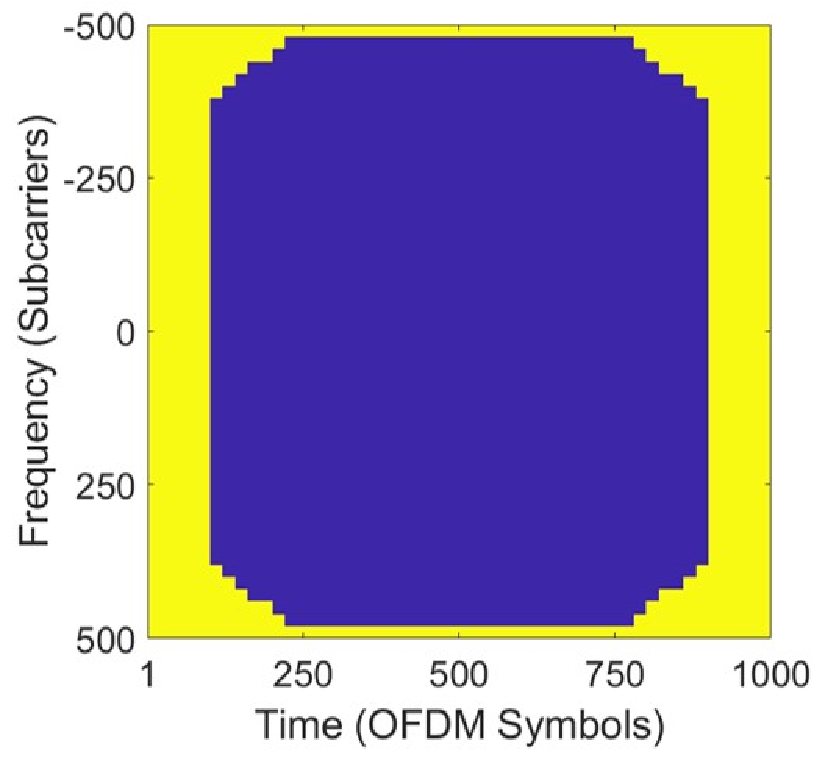}\label{fig:d075}}
    \subfloat[${\epsilon}_{\tau} = 0,{\epsilon}_{\nu}= 1$]{\includegraphics[width=0.5\columnwidth]{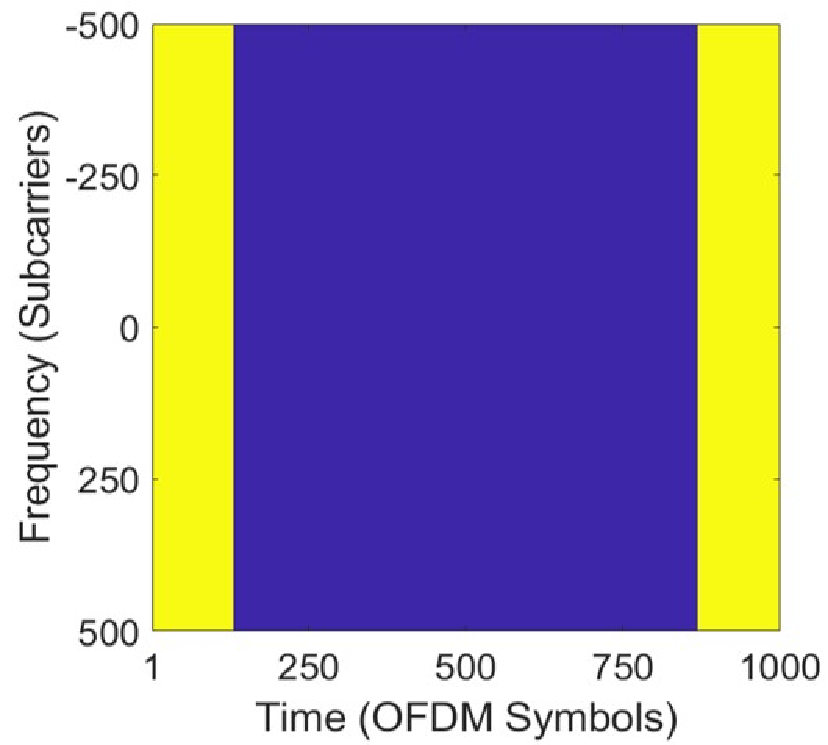}\label{fig:d}}
 \caption{Optimized waveform varying the ROF $\mu$ and weights ${\epsilon}_{\tau}$ and ${\epsilon}_{\nu}$, (a) $\mu  = 0.25, {\epsilon}_{\tau} = {\epsilon}_{\nu}= 0.5$, (b) $\mu  = 0.50, {\epsilon}_{\tau} = {\epsilon}_{\nu}= 0.5$, (c) $\mu = 0.25$, ${\epsilon}_{\tau} = 0.25,{\epsilon}_{\nu}= 0.75$ and (d) $\mu = 0.25$, ${\epsilon}_{\tau} =0, {\epsilon}_{\nu}= 1$. Yellow denotes allocated resources, while blue denotes empty resource bins. }
 \label{fig:waveformb}
\end{figure}

This section focuses on the time-frequency waveform design aimed at minimizing the CRBs for closely placed targets, taking into account practical levels of ROF. In particular, the ISAC waveform design aims to allocate a limited amount of time and frequency resources ($\mu < 1$) to minimize a weighted average of delay and Doppler CRBs while maintaining the desired communication QoS for every UE. Primarily focused on the $K=2$ case, the procedure is generalizable. Resource allocation considers fixed per-resource energy, with potential extensions discussed in Section \ref{sect:TFP_scheduling}. Time-frequency resources for each UE are denoted as $\mathbf{a}_k \in \mathbb{B}^{L \times 1}$ such that
\begin{equation}\label{eq:allocation_vector}
    [\mathbf{a}_k]_{\ell} = \begin{cases} 1,\,\,\text{if the $\ell$th resource is chosen}, \\
    0,\,\,\text{otherwise}.
    \end{cases}
\end{equation}
and $\mathbf{\Sigma} = \sigma \mathbf{1}_M \mathbf{1}_N^T $, thus the transmitted signal can be rewritten as
\begin{equation}\label{eq:designed_waveform}
    \mathbf{X} = \sigma \mathbf{1}_M \mathbf{1}_N^T  \odot \mathbf{S} \odot \mathbf{A}.
\end{equation} 
 $\mathbf{A} = \mathrm{vec}^{-1}\left(\sum_{k=1}^K \mathbf{a}_k\right)$ is the matrix of allocated resources over frequency and time, thus $\mathbf{e}_k = \sigma^2 \mathbf{a}_k$. The overall Tx energy is $E = \sigma^2 \| \mathbf{A} \|_F^2 $.
The waveform design problem, solely concerning the selection of time-frequency resources under limited resource occupancy $\mu$, can be formulated as in \cite{mura2023waveform}:
\begin{subequations}\label{eq:optProb1}
\begin{alignat}{2} 
&\underset{\{\mathbf{a}_k\}_{k=1}^K}{\mathrm{minimize}}  &\quad&  \epsilon_{\tau} \,\mathrm{tr}\left(\frac{\mathbf{C}_{\tau}(\mathbf{a}_k)}{\Delta \tau^2}\right) + \epsilon_{\nu}\mathrm{tr}\,\left(\frac{\mathbf{C}_{\nu}(\mathbf{a}_k)}{\Delta \nu^2}\right)\\
&\mathrm{s.\,t.} &      & \frac{1}{L}\sum_{\ell=1}^L\log_2(1+{\gamma}_{k,\ell})\geq  \overline{\eta},\,\,\, \forall k, \label{eq:prob1_constraint1}\\
&  &      & \sum_{k=1}^K [\mathbf{a}_k]_{\ell} \leq 1,\,\,\, \forall \ell, 
\label{eq:prob1_constraint2}\\
&  &      & \sum_{k=1}^K \mathbf{1}^T\mathbf{a}_k  \leq \mu L, \,\,\label{eq:prob1_constraint3}\\
%&  &      & [\mathbf{x}_k]_{\ell} \in \{0, 1\},\forall i,\forall k\label{eq:prob1_constraint4}\\
&  &      & 1 \leq \ell \leq L, \nonumber\\
&  &      & 1 \leq k \leq K \nonumber.
\end{alignat}
\end{subequations}
where $\mathbf{C}_{\tau}(\mathbf{a}_k) \in \mathbb{R}^{K\times K}$ and $\mathbf{C}_{\nu}(\mathbf{a}_k) \in \mathbb{R}^{K\times K}$ denote the CRBs on delay and Doppler estimation, respectively, that exhibit a non-linear dependence w.r.t. the allocated resources $\mathbf{a}_k$, as detailed in Appendix \ref{app:CRB} for $K=2$. 
The cost function quasi normalizes CRBs by maximum delay and Doppler resolutions ($\Delta_{\tau}=1/B$ and $\Delta_{\nu}=1/(NT)$) to ensure uniformity and introduces dimensionless weights ($\epsilon_{\tau}$ and $\epsilon_{\nu}$). These weights influence the trade-off between delay and Doppler CRBs in optimization. The objective in \eqref{eq:optProb1} minimizes the combined CRBs adjusted by $\epsilon_{\tau}$ and $\epsilon_{\nu}$. The constraint in \eqref{eq:prob1_constraint1} establishes the QoS requirement, expressed through a threshold on the achievable rate $\overline{\eta}$ in [bits/s/Hz] across all UEs. The communication SNR related to the $k$th UE and $\ell$th resource is 
\begin{align}
{\gamma}_{k,\ell}= \frac{\sigma^2\,\,[\mathbf{a}_k]_\ell |H_k|^2 }{N_0}.
\end{align}

The achievable rate constraint ties resource quantity to UE selection, maintaining constant energy allocation per resource. Variations in target distances determine resource allocation, with targets experiencing higher pathloss assigned more resources. Constraint \eqref{eq:prob1_constraint2} mitigates multi-user interference by assigning each resource to a single UE. Occupancy constraint \eqref{eq:prob1_constraint3} limits total resource allocation, with $\mu<1$ defining the maximum allowable fraction.

The optimization problem \eqref{eq:optProb1} is non-convex, posing challenges for solution. Through auxiliary variables detailed in Appendix \ref{app:OptimizationCRB}, convexity can be achieved. This results in a mixed-integer conic programming problem (MICP), given the binary nature of the model. MICP involves continuous and discrete variables, demanding computational resources for branch-and-cut (BnC) methods. To limit the BnC algorithm complexity, time and frequency resources are grouped into subchannels and time slots, which aligns with the 3GPP standard. \cite{rahman20215g}.

Figures \ref{fig:waveformb} show waveform design across time and frequency domains. Figs. \ref{fig:waveformb1} and \ref{fig:waveformb2} depict resource allocation influenced by $\mu$, with equal $\epsilon_{\tau}$ = $\epsilon_{\nu}$ = $0.5$. The time-frequency resources are allocated at the grid boundaries to minimize CRB while adhering to communication constraints. Conversely, Fig. \ref{fig:d075} prioritizes minimizing Doppler CRB ($\epsilon_{\tau} = 0.25$, $\epsilon_{\nu} = 0.75$), favoring frequency axis resources. Extreme condition ($\epsilon_{\tau}$ = $0$, $\epsilon_{\nu}$ = $1$) in Fig. \ref{fig:d} exclusively minimizes Doppler CRB, allocating all resources along the frequency axis.

\section{Joint Time-Frequency-Energy Optimization}\label{sect:TFP_scheduling}

While the previous section focused solely on waveform design in the time-frequency domain, this section introduces an additional variable: the allocated energy per resource. Consequently, unlike \eqref{eq:optProb1}, the ability to vary the allocated energy across time and frequency enhances the degrees of freedom available for waveform design. Therefore, in the following, we extend \eqref{eq:optProb1} to account for a joint time-frequency-energy allocation. Now, $\mathbf{e}_k \in \mathbb{R}^{L \times 1}$ denotes the per-resource energy allocated to the $k$th UE, while $\mathbf{a}_k \in \mathbb{B}^{L \times 1}$ denotes the Boolean time-frequency allocation vector. The joint time-frequency-power optimization is formulated as follows: 
\begin{subequations}\label{eq:optProb2}
\begin{alignat}{2} 
&\underset{\{\mathbf{e}_k,\mathbf{a}_k\}_{k=1}^K}{\mathrm{minimize}}  &\quad&  \epsilon_{\tau} \,\mathrm{tr}\left(\frac{\mathbf{C}_{\tau}}{\Delta \tau^2}\right) + \epsilon_{\nu}\mathrm{tr}\,\left(\frac{\mathbf{C}_{\nu}}{\Delta \nu^2}\right)\\
&\mathrm{s.\,t.} &      & \frac{1}{L}\sum_{\ell=1}^{L}\log_2(1+{\gamma}_{k,\ell})\geq  \overline{\eta},\,\,\, \forall k, \label{eq:prob2_constraint1}\\
&  &      & \sum_{k=1}^K [\mathbf{a}_k]_{\ell} \leq 1,\,\,\, \forall \ell, 
\label{eq:prob2_constraint2}\\
&  &      & \sum_{k=1}^K \mathbf{1}^T \mathbf{a}_k  \leq \mu L, \,\,\label{eq:prob2_constraint3}\\
%&  &      & [\mathbf{x}_k]_{\ell} \in \{0, 1\},\forall i,\forall k\label{eq:prob1_constraint4}\\
&  &      &\sum_{k=1}^K \mathbf{1}^T \mathbf{e}_k \leq E_{\mathrm{max}},\label{eq:probp2_constraint4}\\
&  &      &0 \leq [\mathbf{e}_k]_{\ell} \leq [\mathbf{a}_k]_{\ell}\,\sigma_{\mathrm{max}}^2, \,\, \forall \ell,\forall k,\label{eq:probp2_constraint5}\\
&  &     & \big \lvert [\mathbf{e}_k]_{\ell}-[\mathbf{e}_k]_{\ell+1}\big\rvert  \leq \Delta T,\ \forall \ell, \forall k,\label{eq:probp2_constraint6}\\
&  &      & 1 \leq \ell \leq L, \nonumber\\
&  &      & 1 \leq k \leq K \nonumber.
\end{alignat}
\end{subequations}
where the dependence of the CRBs from the optimization variables $\{\mathbf{e}_k,\mathbf{a}_k\}_{k=1}^K$ is omitted for simplicity. Constraints \eqref{eq:prob2_constraint1}-\eqref{eq:prob2_constraint3} are similar to \eqref{eq:prob1_constraint1}-\eqref{eq:prob1_constraint3}, except that the QoS involves a non-constant energy in the SNR
\begin{equation}
    \gamma_{k,\ell} = \frac{[\mathbf{e}_k \odot \mathbf{a}_k]_\ell \, |H_k|^2}{N_0}.
\end{equation}
Constraint \eqref{eq:probp2_constraint4} expresses the overall energy budget (on the whole time-frequency grid), while \eqref{eq:probp2_constraint5} establishes a relationship between the two sets of optimization variables, ensuring that energy allocation per resource $[\mathbf{e}_k]_\ell > 0$ only occurs when the resource is actively allocated ($[\mathbf{a}_k]_\ell = 1$). 
In this context, $\sigma_{max}^2$ represents the maximum energy per time-frequency resource. Constraint \eqref{eq:probp2_constraint6} regulates energy smoothness, limiting energy gradient across time and frequency to $\Delta$. Including \eqref{eq:probp2_constraint6} aims to reduce sidelobes in the ambiguity function of the ISAC waveform and meets requirements for gradual Tx power changes in power amplifiers. The choice of $\Delta$ impacts estimation accuracy and QoS elaborated in Section \ref{sect:numerical_results}.

\begin{figure}[!t]
    \centering
    \subfloat[][$\Delta  = -30$ dB]{ \includegraphics[width=0.5\columnwidth]{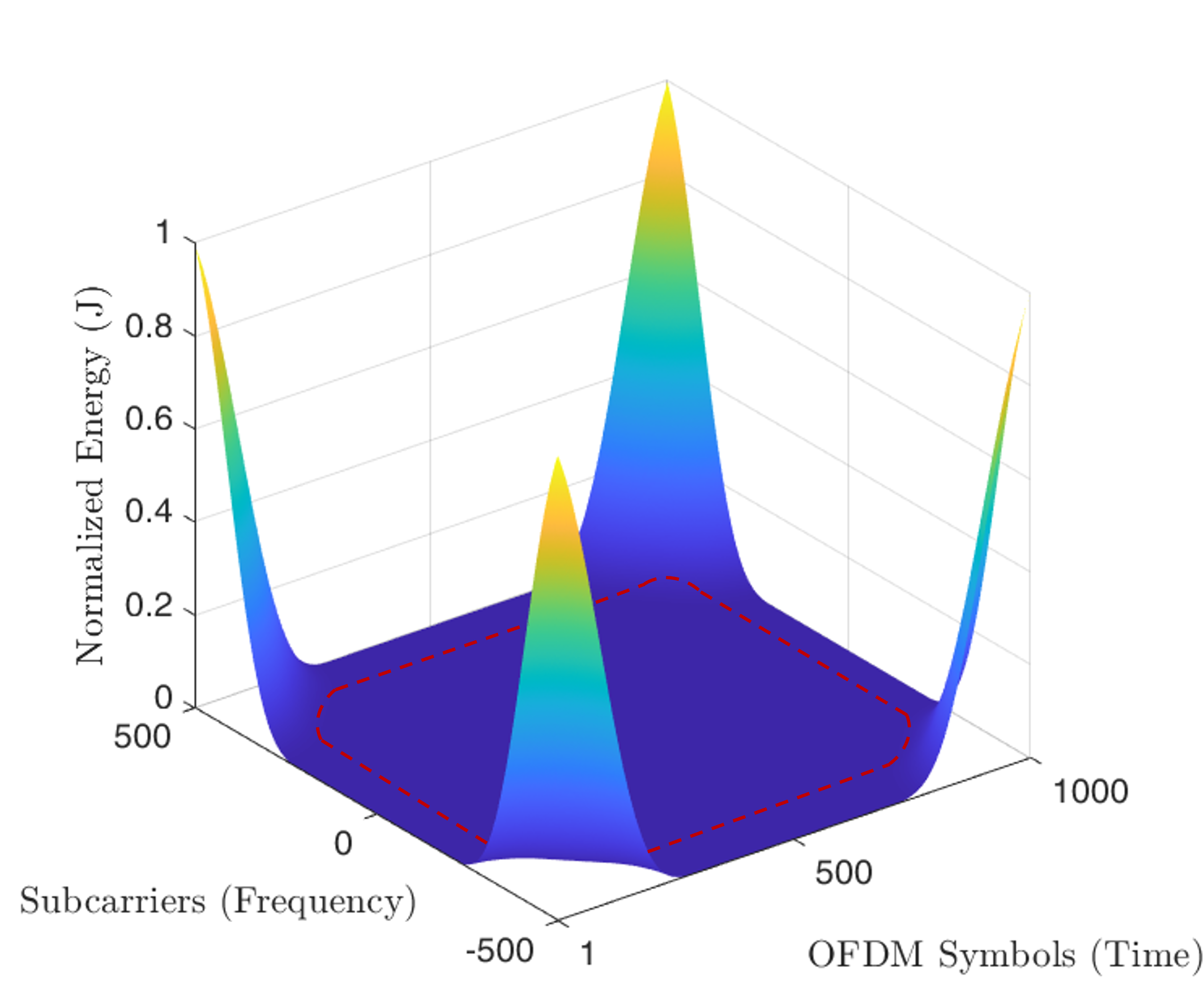}\label{fig:waveformg1}}\subfloat[ ][$\Delta  = -10$ dB]{\includegraphics[width=0.5\columnwidth]{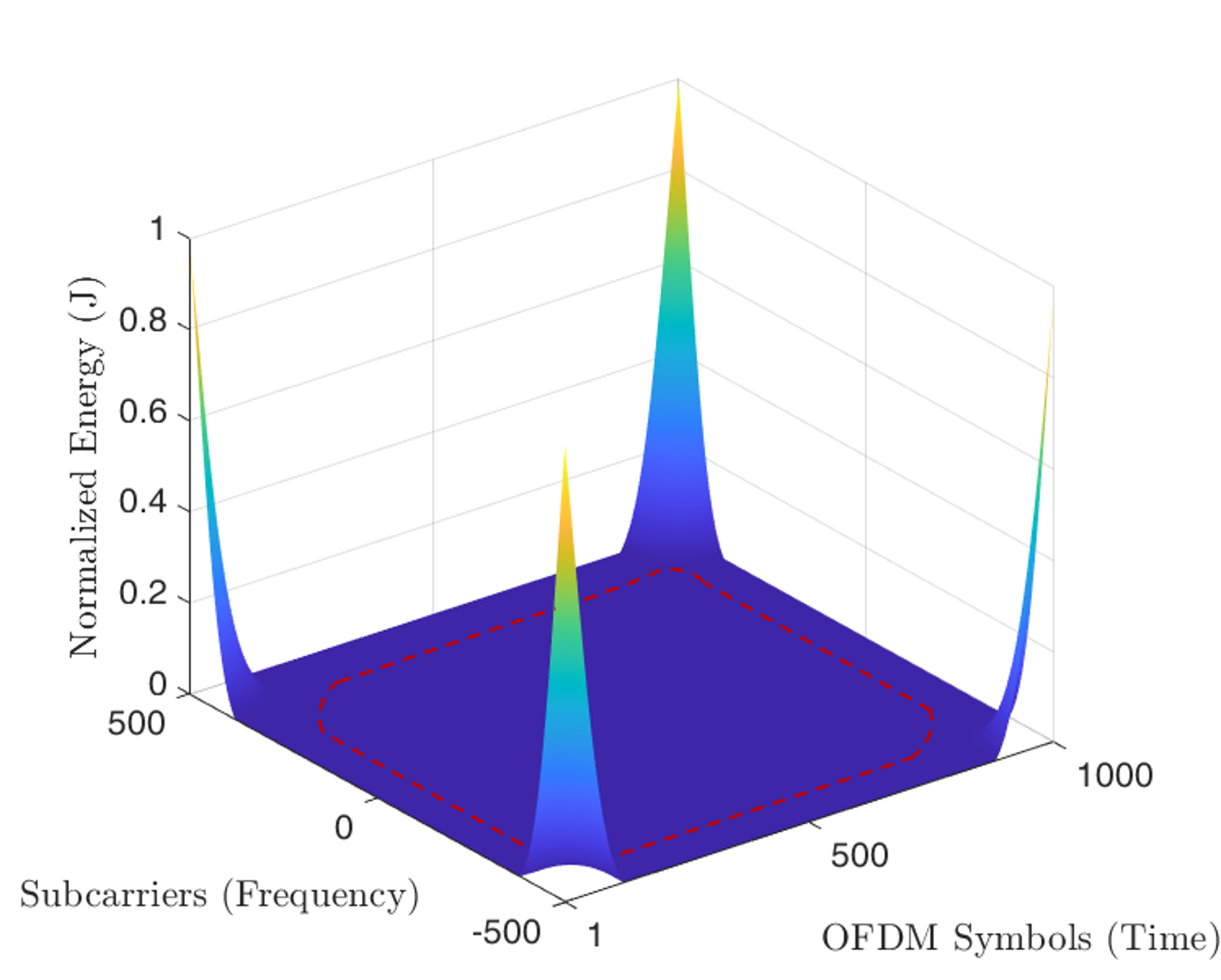}\label{fig:waveformg2}}\\
    \subfloat[][${\epsilon}_{\tau} = 0.25,{\epsilon}_{\nu}= 0.75$]{ \includegraphics[width=0.5\columnwidth]{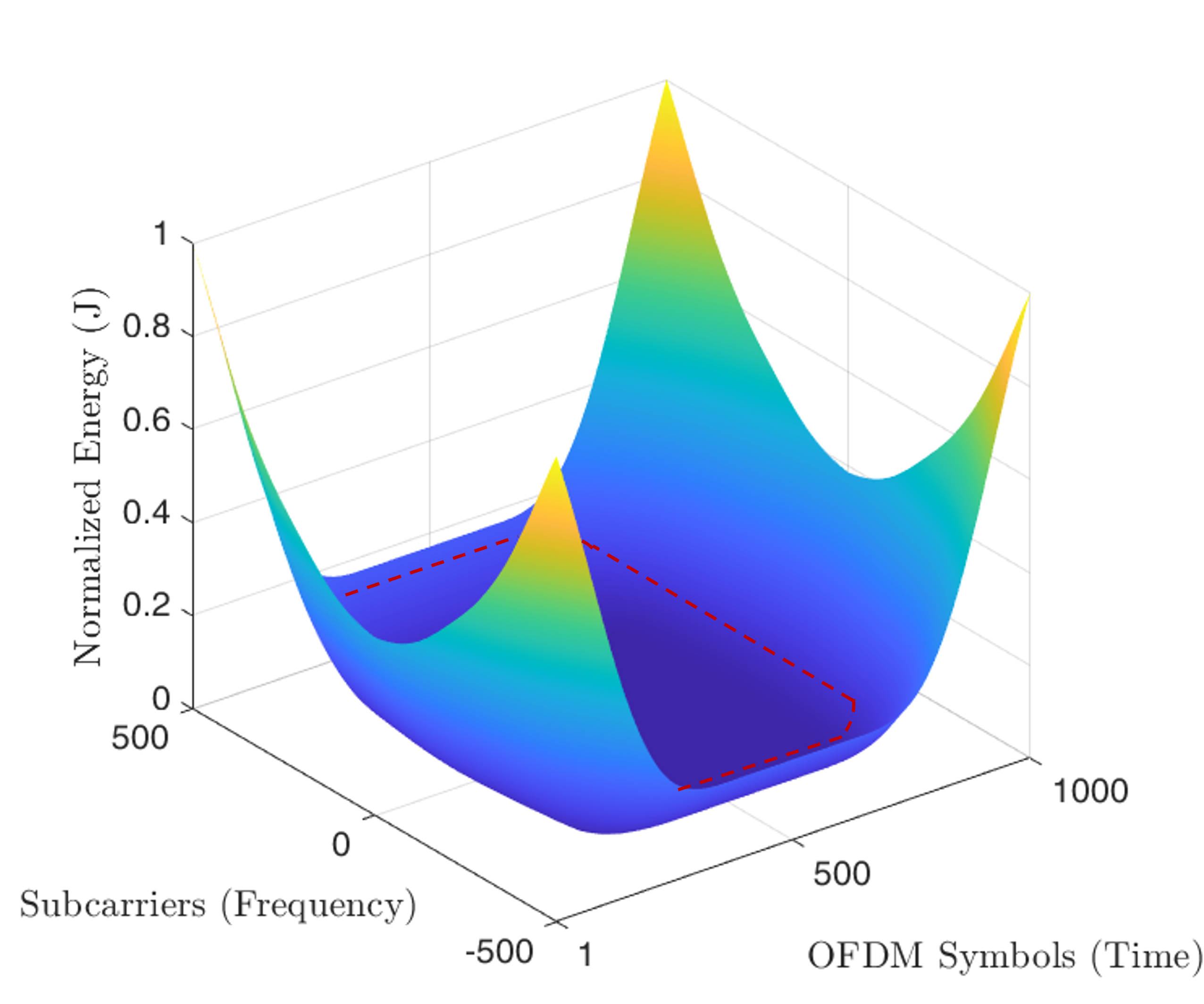}\label{fig:g075}}
    \subfloat[ ][${\epsilon}_{\tau} = 0,{\epsilon}_{\nu}= 1$]{\includegraphics[width=0.5\columnwidth]{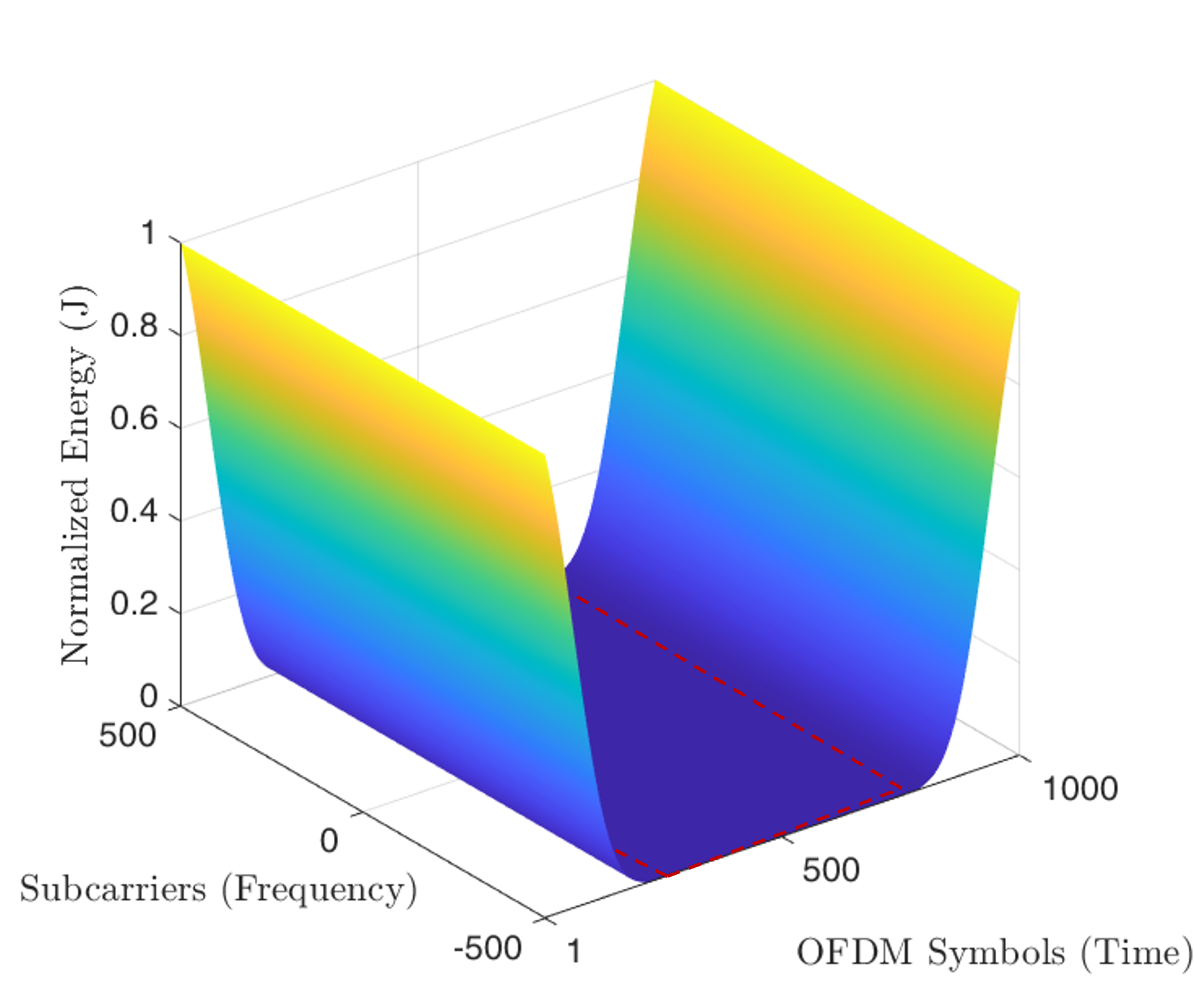}\label{fig:g}}
    \caption{Optimized waveform for $\mu = 0.25$, (a) ${\epsilon}_{\tau} = {\epsilon}_{\nu}= 0.5$, $\Delta = -30$ dB, (b) ${\epsilon}_{\tau} = {\epsilon}_{\nu}= 0.5$, $\Delta = -10$ dB, (c) ${\epsilon}_{\tau} = 0.25,{\epsilon}_{\nu}= 0.75, \Delta = -30$ dB, and (d) ${\epsilon}_{\tau} = 0,{\epsilon}_{\nu}= 1, \Delta = -30$ dB. quasi normalized energy: {$\mathrm{max} (\mathbf{e}) = 1$ J}. The red dashed line represents the boundary between allocated resources ( at the borders) and empty resources (at the center).}  
    \label{fig:waveformg}
\end{figure}

The problem in \eqref{eq:optProb2} remains an MICP. Unlike the problem in \eqref{eq:optProb1}, it includes the linear constraints of \eqref{eq:probp2_constraint4}-\eqref{eq:probp2_constraint6}. However, incorporating these constraints does not significantly affect the problem solver or its complexity. We can still tackle it using BnB by adjusting the granularity of the resources to be allocated and appropriately scaling the linear constraints in \eqref{eq:probp2_constraint4}-\eqref{eq:probp2_constraint6}.

Figure \ref{fig:waveformg} showcases the waveform design spanning energy, time, and frequency domains, with Fig.\ref{fig:waveformg1} and \ref{fig:waveformg2} depicting the influence of maximum energy gradient $\Delta$ while maintaining balanced weights $\epsilon_\tau = \epsilon_\nu = 0.5$. The objective is to minimize the CRB while adhering to communication constraints and energy gradient constraints, favoring higher power levels towards frequency-time grid edges, guided by $\Delta$. Conversely, Fig. \ref{fig:g075} prioritizes minimizing the Doppler CRB over the delay CRB ($\epsilon_{\tau} = 0.25$, $\epsilon_{\nu} = 0.75$), amplifying energy levels along the frequency axis. Fig. \ref{fig:g} depicts an extreme condition emphasizing Doppler CRB reduction ($\epsilon_{\tau} = 0$, $\epsilon_{\nu} = 1$), allocating high energy levels exclusively along the frequency axis, with extremely low energy along the time axis. 

\section{Sensing Channel Interpolation and Parameters Estimation}\label{sect:interp}

This section introduces a novel framework for sensing channel estimation and interpolation, effectively addressing the issue of high sidelobe levels resulting from the low ROF.
After the waveform design over time, frequency, and possibly energy, the ISAC BS estimates the delay and Doppler shifts of the $K$ UEs/targets.
The estimation process is grounded in the maximum likelihood (ML) framework. This methodology involves leveraging the received signal \eqref{eq:vec_rxsignal}, to formulate the ML estimation as
\begin{equation} (\widehat{\boldsymbol{\tau}},\widehat{\boldsymbol{\nu}}) = \underset{\boldsymbol{\tau},\boldsymbol{\nu} }{\mathrm{argmin}} \left(\left\|\mathbf{r} - \mathbf{x}\odot \mathbf{h}_s(\boldsymbol{\tau},\boldsymbol{\nu} )\right\|^2_2\right),
\end{equation}
where vectors $\boldsymbol{\tau} = [\tau_1,...,\tau_K]^T$ and $\boldsymbol{\nu} = [\nu_1,...,\nu_K]^T$ represent the delays and Doppler shifts to be determined. While ML estimation could involve an exhaustive search across the delay and Doppler shift domains, practical systems employ a suboptimal approach through three sequential steps: $(i)$ estimate the sensing channel $\mathbf{H}_s$,
$(ii)$ employ an DFT-IDFT transform to map the estimated sensing channel $\mathbf{H}_s$ from the frequency-time domain to the delay-Doppler domain and $(iii)$ estimate $(\widehat{\boldsymbol{\tau}},\widehat{\boldsymbol{\nu}})$ via peak searching.

The time-frequency sensing channel matrix $\mathbf{H}_s$ is first estimated by the least squares (LS) approach over the allocated resources as:
\begin{align}\label{eq:estimated_channel}
    [\widehat{\mathbf{H}}_s]_{mn} = \begin{dcases}
        [\mathbf{R}]_{mn} \odot [\mathbf{X}]^{-1}_{mn} & \text{for $[\mathbf{A}]_{mn}=1$}\\
        0 &\text{for $[\mathbf{A}]_{mn}=0$}
    \end{dcases}
\end{align}
where $\mathbf{X}$ is the waveform by solving either \eqref{eq:optProb1} or \eqref{eq:optProb2}. The channel is subsequently mapped into the delay-Doppler domain using a DFT-IDFT pair, outlined as follows: 
\begin{equation}\label{eq:sensing_channel_DD}
   {\widehat{\mathbf{H}}}_s^{\mathrm{DD}} = \mathbf{\Theta}_M \widehat{\mathbf{H}}_s \mathbf{\Theta}_N^H
\end{equation}
where $\mathbf{\Theta}_M\in\mathbb{C}^{M\times M}$ and $\mathbf{\Theta}_N\in\mathbb{C}^{N\times N}$ are DFT matrices such that $\|\mathbf{\Theta}_M\|_F=\sqrt{M}$ and $\|\mathbf{\Theta}_N\|_F=\sqrt{N}$. 

For full resource occupancy($\mu=1$), the delay-Doppler sensing channel matrix exhibits a linear combination of scaled and shifted sinc functions (the expression is reported in \eqref{eq:discrete_DD_sensing_channel}), narrowing with resolution. However, for $\mu < 1$, the sensing channel's expression differs markedly, displaying higher sidelobes due to unused resources, impacting target discrimination and resolution. Ghost peaks in the channel matrix are influenced by resource allocation, affecting the system's effective resolution, especially for closely spaced or coupled targets. Consequently, the periodogram approach becomes impractical\footnote{The same considerations can be drawn by inspection of the ambiguity function of the Tx signal $\mathbf{X}$}.
  
\begin{figure*}[!t]
\begin{equation}\label{eq:discrete_DD_sensing_channel}
\begin{split}
[{\widehat{\mathbf{H}}}_s^{\mathrm{DD}}]_{ij} =  \frac{1}{\sqrt{MN}}\sum_{k=1}^{K} \beta_{k} e^{- j 2 \pi \nu_k \tau_k}\, \frac{\sin\left(\pi(j - \nu_{k}/\Delta \nu)\right)}{\sin\left(\frac{\pi}{N}(j - \nu_{k}/\Delta \nu)\right)}\, \frac{\sin\left(\pi(i - \tau_{k}/\Delta \tau)\right)}{\sin\left(\frac{\pi}{M}(i - \tau_{k}/\Delta \tau)\right)}.
\end{split}
\end{equation}
\hrulefill
\end{figure*}

To mitigate this issue in the case of $\mu < 1$, we propose a technique involving the sensing channel interpolation across frequency and time, employing matrix completion. The formulation of the matrix completion problem is as follows \cite{kalogerias2013matrix}:
\begin{subequations}
\begin{alignat}{2} 
&\underset{{\mathbf{H}_s}}{\mathrm{minimize}}  &\quad& \mathrm{rank}({\mathbf{H}_s})\label{eq:optkp1Prob_interp}\\
&\mathrm{s.\,t.} &  &[{\mathbf{H}_s}]_{mn} =  [\widehat{\mathbf{H}}_s]_{mn} \,\,\, \text{for $[\mathbf{A}]_{nm}=1$}\label{eq:interpconstraint}.
\end{alignat}
\end{subequations}
The linear constraint in \eqref{eq:interpconstraint} imposes that the entries of the optimization variable $\mathbf{H}_s$ must be equal to the entries of the estimated sensing time-frequency channel $\widehat{\mathbf{H}}_s$. 
The endeavor to minimize the matrix rank while complying with linear (affine) constraints represents an NP-hard challenge frequently tackled via nuclear quasi norm minimization. Despite its convexity, this method offers a too approximate solution for channel rank estimation, making it inadequate for the ROFs detailed in this paper. This inadequacy arises due to its demand for a greater number of samples to facilitate interpolation, as highlighted in \cite{malek2015performance, marjanovic2012l_q}.

Alternatively, problem \eqref{eq:optkp1Prob_interp} can be approached by relaxing the objective function using the Schatten $p$-quasi norm, defined as:
\begin{equation}\label{eq:schatten}
\|\mathbf{H}_s\|_p^p = \left(\sum_{r}^{\mathrm{min}(M,N)} \lambda_r^p\right)^{\frac{1}{p}}
\end{equation}
where $p \in (0,1]$, and $\lambda_r$ represents the $r$-th singular value of $\mathbf{H}_s$. The $p$-Schatten quasi norm presents a flexible balance between convexity and rank approximation within matrix optimization. By setting $p = 1$, the Schatten quasi norm reduces to the nuclear quasi norm (sum of eigenvalues) resulting in a convex problem, while $p \rightarrow 0$ provides a more accurate estimator of matrix rank and results in a non-convex problem. The $p$-value allows adjustment between computational complexity and recovery accuracy in problems concerning matrix rank minimization. Consequently, the matrix completion problem can be reformulated as:
\begin{subequations}\label{eq: interp}
\begin{alignat}{2} 
&\underset{{\mathbf{H}_s}}{\mathrm{minimize}}  &\quad& \|\mathbf{H}_s\|_p^p\label{eq:optkp1Prob_interp1}\\
&\mathrm{s.\,t.} &  &||[{\mathbf{H}_s}]_{\mathbf{A}}-[\widehat{\mathbf{H}}_s]_{\mathbf{A}}||_2 \leq \epsilon\,\,\,\label{eq:interpconstraint1}.
\end{alignat}
\end{subequations}
with $\epsilon$ is a small positive constant and $[\widehat{\mathbf{H}}_s]_{\mathbf{A}}||_2$ are the channel samples, defined where $[\mathbf{A}]_{mn} =1$. We denote as $\widetilde{\mathbf{H}}_s$ the solution of the optimization problem in \eqref{eq:optkp1Prob_interp1} and in the further subsections the conditions for solving the problem are discussed. 
\subsection{Conditions for solving \eqref{eq: interp}}
\begin{figure}[t!]    \includegraphics[width=0.65\columnwidth]{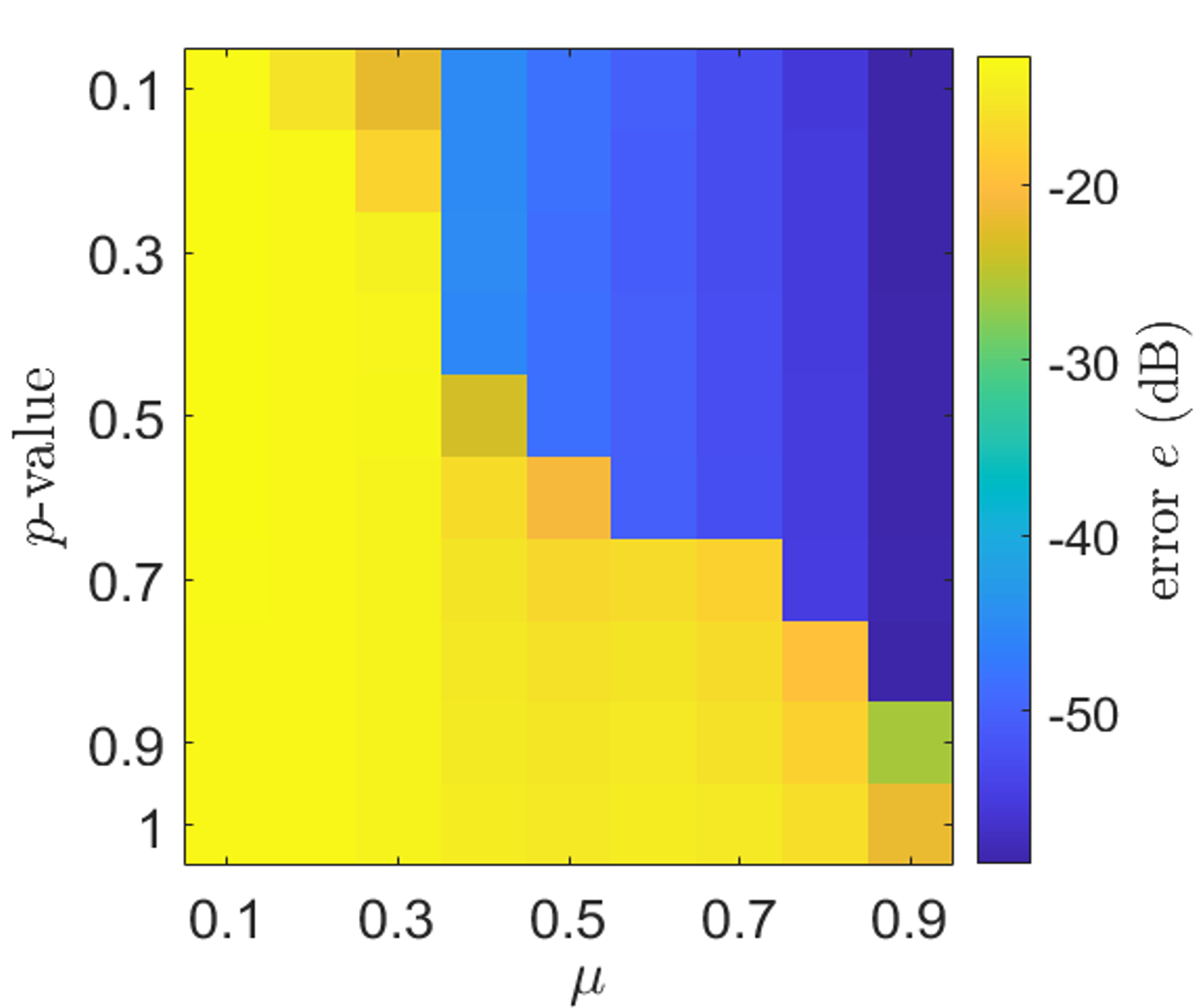} \label{fig:error_mu_p}\vspace{-0.2cm}
    \caption{Matrix completion error in dB versus the $p$-value of the Schatten quasi norm and $\mu$}
\end{figure}

This section provides the conditions for solving the  Schatten $p$-quasi norm matrix completion problem. 
Despite the non-convex nature of problem \eqref{eq: interp}, numerically efficient algorithms have been proposed in \cite{8931253,marjanovic2012l_q}. The algorithm used for solving \eqref{eq: interp} involves non-convex matrix completion via the iterative singular value thresholding algorithm (ISTVA) in \cite{8931253}.
A proper selection of the parameter $p$ allows trading between the capacity to effectively recover the sensing channel rank (for $p\rightarrow 0$) and the computational complexity (lower for $p\rightarrow 1$). The matrix completion error $e  = ||\mathbf{H}_s-\widetilde{\mathbf{H}}_s||_2$ is computed by evaluating the quasi norm between the real sensing channel $\mathbf{H}_s$ and the solution $\widetilde{\mathbf{H}}_s$ of the problem in \eqref{eq:optk2Prob_r} by varying the resource occupancy parameter $\mu$ and the $p$-value, as depicted in Fig.\ref{fig:error_mu_p}. 
 A low value of $p$ enables a better reconstruction of the matrix, but it may suffer from relatively higher computational complexity because
of the singular value decomposition (SVD) computed
at each iteration and more iteration steps to reach the stopping
criteria\cite{8931253}. 
Empirical simulations indicate that an error of around $-25$ dB yields satisfactory interpolation performance when noise is not considered. Thus, with $\mu$ values between 0.2 and 0.4 and a p-value of 0.1, the algorithm can achieve effective interpolation.
A recognized criterion ensuring the correct retrieval of the sensing channel matrix $\mathbf{H}_s$ with minimum rank and in case of deterministic sampling is the satisfaction of the \textit{isomeric condition} and \textit{relative well conditionedness}\cite{liu2019matrix}.  
\begin{figure}\label{fig:cond_number}
    \centering
    \includegraphics[width=0.9\columnwidth]{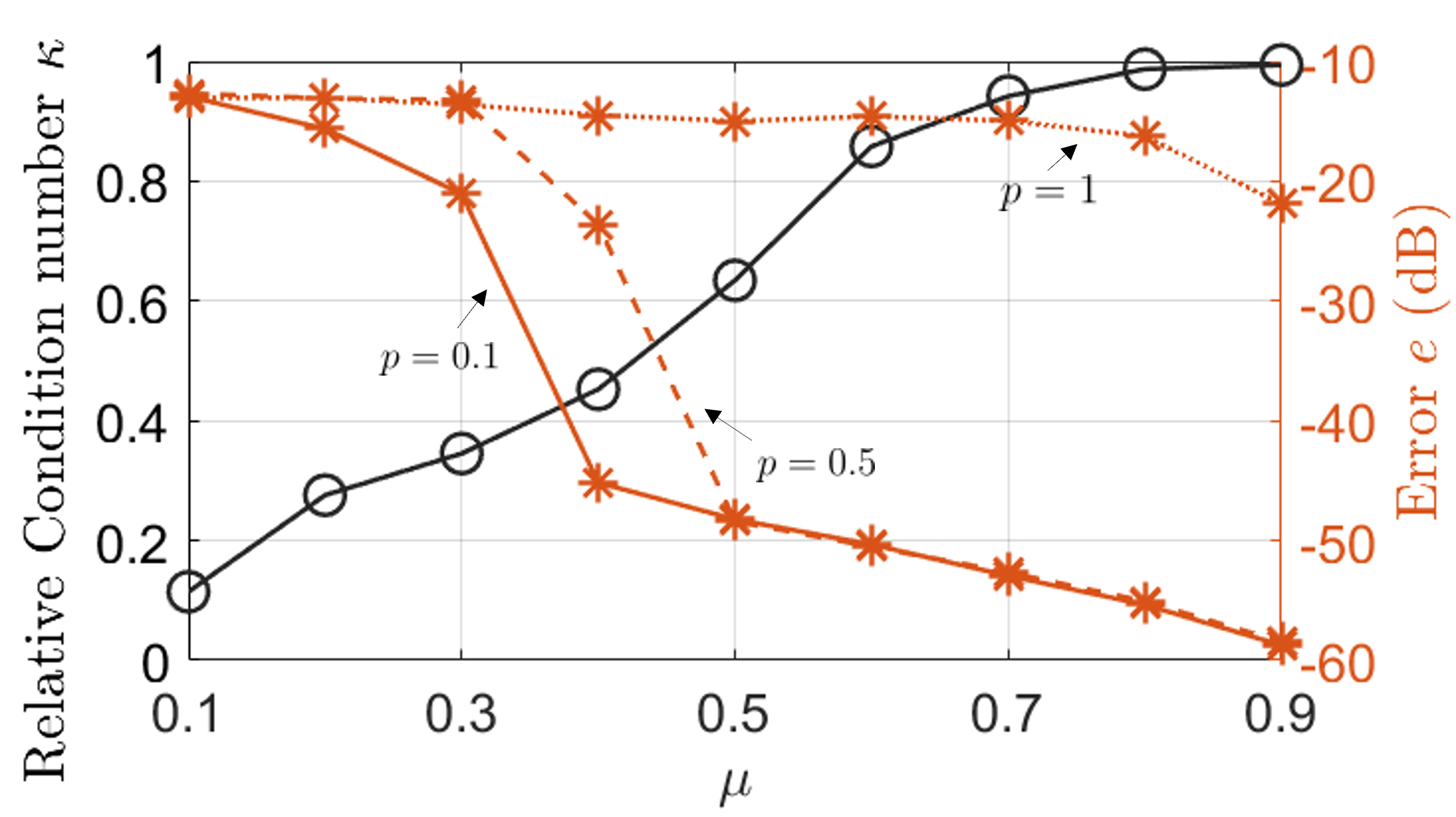}\vspace{-0.2cm}
    \caption{Relative condition number $\kappa$ and error $e$ in dB versus $\mu$}
\end{figure}

The isomeric condition interlaces the rank and the coherence of the matrix $\mathbf{H}_s$ with the specific locations and quantity of the observed entries. In particular, the matrix $\widehat{\mathbf{H}}_s$ is called $\mathbf{A}$-\textit{isomeric} if the submatrix related to the sampling $\mathbf{A}\subseteq \{1,...,M\} \times \{1,...,N\}$ is defined such that
\begin{align}
    \mathrm{rank}([\widehat{\mathbf{H}}_s]_{\mathbf{A}}) = \mathrm{rank}(\mathbf{H}_s),
\end{align}
Moreover, the matrix $\widehat{\mathbf{H}}_s$ is called $\mathbf{A}/\mathbf{A}^T$-isomeric if $\widehat{\mathbf{H}}_s$ is $\mathbf{A}$-isomeric and $\widehat{\mathbf{H}}_s^T$ is $\mathbf{A}^T$-isomeric. To solve the matrix completion problem in \eqref{eq:optProb1} it is necessary to verify that $\widehat{\mathbf{H}}_s$ is $\mathbf{A}/\mathbf{A}^T$-isomeric. Whenever this isomeric condition is violated, there exist infinitely many solution matrices that can fit the observed entries. 

In general, isomerism typically ensures that the sampled sub-matrices $[\widehat{\mathbf{H}}_s]_{\mathbf{A}}$ and $[\widehat{\mathbf{H}}_s]_{\mathbf{A}}^T$ are not rank-deficient, but there is no guarantee that these sub-matrices are well-conditioned. To compensate for this weakness, the hypothesis of \textit{relative well
conditionedness}, which encourages the smallest singular value of the sampled sub-matrices to be far from 0, must be satisfied. In particular, the \textit{$\mathbf{A}/\mathbf{A}^T$-relative condition number} is defined as in \cite{liu2019matrix}
\begin{align}
    \kappa = \mathrm{min}( \kappa_{\mathbf{A}}, \kappa_{\mathbf{A}^T})
\end{align}
with $\kappa_{\mathbf{A}} = \|[\widehat{\mathbf{H}}_s]_{\mathbf{A}}(\mathbf{H}_s)^{\dag}\|^2$ 
measuring how much information of a matrix $\mathbf{H}_s$ is contained in the sampled sub-matrix $[\widehat{\mathbf{H}}_s]_{\mathbf{A}}$. The condition number $\kappa_{\mathbf{A}^T}$ is computed in the same way by considering the matrix $\mathbf{H}_s^T$. To ensure that the matrix $\mathbf{H}_s$ is recoverable, $ \kappa$ must be sufficiently high, thus ensuring the \textit{relative well-conditionedness} property. Figure \ref{fig:cond_number} illustrates the attained relative condition number relative to the resource occupancy parameter $\mu$. As the condition number increases sufficiently, the algorithm demonstrates improved performance. Specifically, when $p = 0.1$, the associated condition number required to achieve an error $e$ smaller than -20 dB is approximately 0.3, whereas it becomes higher when $p = 0.5$.

\begin{figure}[t!]
 \centering
 \includegraphics[width=0.9\columnwidth]{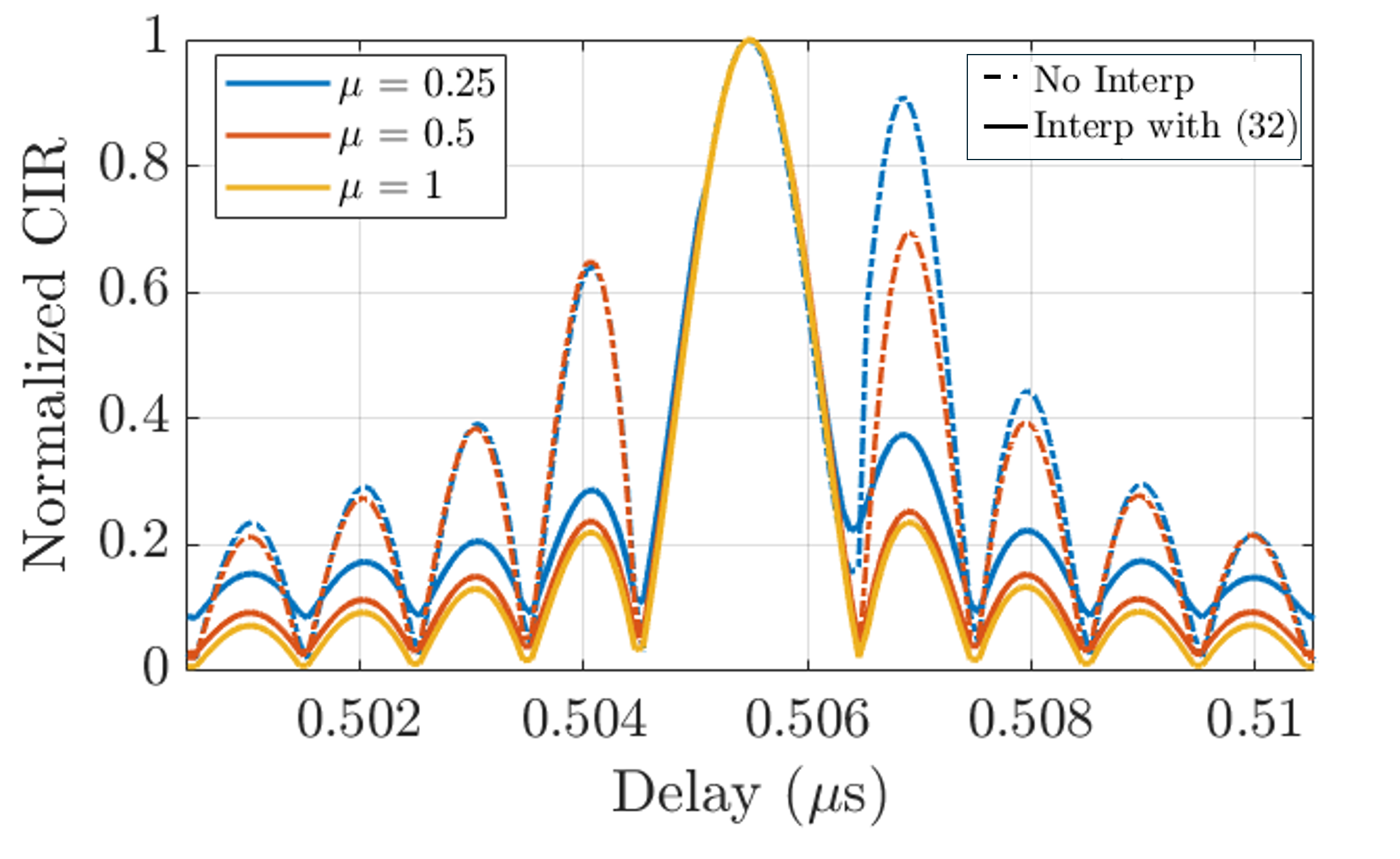}
 \caption{Comparison of the CIR before and after interpolation for different ROF with the full grid occupancy ($\mu$ = 1)}
 \label{fig:cir}
\end{figure}

\subsection{Example of channel interpolation}
In this subsection, we demonstrate channel interpolation using varying resource occupancy ratios 
$\mu$. The interpolation relies on the optimized waveform discussed in Sect \ref{sect:interp}. 
Figure \ref{fig:cir} provides empirical evidence that the suggested interpolation method effectively diminishes sidelobes when examining the estimated quasi normalized sensing channel impulse response (CIR). The CIR is displayed along the delay axis both before and after employing matrix completion interpolation. At a low ROF ($\mu$ = 0.25), the sidelobe level becomes notably prominent, potentially introducing inaccuracies in delay estimations. Through the proposed interpolation method, there is an observed reduction of sidelobe levels by a factor of $2 \times$ under conditions of low resource occupation. This reduction facilitates an impulse response of the channel that approaches the optimal state with complete bandwidth occupancy.
\begin{remark}It is important to remark that resource clustering at bandwidth edges minimizes CRBs, but it negatively impacts on the interpolation performance. However, it is possible to balance the resource distribution and CRB minimization to achieve good interpolation performance, especially with a ROF smaller than the ones analyzed in the paper ($\mu < 0.25$). For instance,  considering that 90\% of the resources are allocated as in \eqref{eq:optProb2}, while the remaining 10\% is periodically distributed across the bandwidth leads to a sidelobe reduction of around $12 \%$ for $\mu = 0.25$ and $26\%$ for $\mu = 0.5$, when no interpolation is performed. When interpolating through \eqref{eq: interp} it achieves the full bandwidth occupation ($\mu =1$) performance. However, this leads to an increase of the CRBs of $3\%$ and $7\%$ when $\mu = 0.25$ and $\mu = 0.5$, respectively. 
\end{remark}

\begin{table}[b!]
    \centering
    \footnotesize
    \caption{Simulation Parameters}
    \begin{tabular}{l|c|c}
    \toprule
        \textbf{Parameter} &  \textbf{Symbol} & \textbf{Value(s)}\\
        \hline
        Carrier frequency & $\Delta f$  & $1$ MHz \\
        Bandwidth & $B$ & $1$ GHz\\
        Size of resource blocks & $N_b$ & $10$ \\
        Number of subcarriers & $M$ & $1000$\\
        Number of symbols & $N$ & $1000$\\
        Symbol duration & $T$ & $1$ $\mu$s \\
        OFDM occupancy ratio & $\mu$ & $0.1-1$ \\
        Range & $R$ & $50$ m\\
        \bottomrule
    \end{tabular}
    \label{tab:SimParam}
\end{table}
\begin{figure}[t!]
    \centering
    \subfloat[][$\mu$ = 0.25]{\includegraphics[width=0.9\columnwidth]{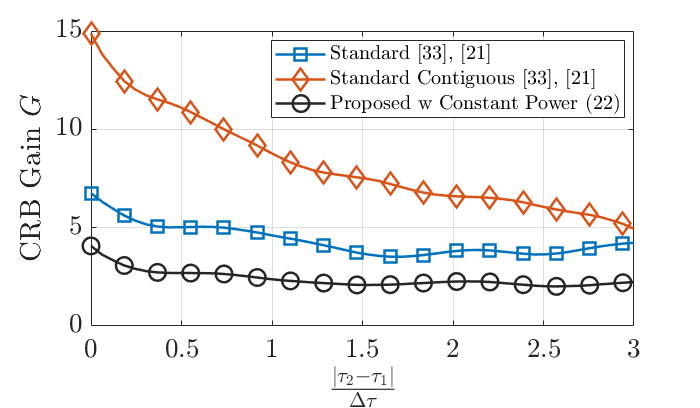}\label{subfig:CRBGain_025}} \\ \vspace{-0.2cm}
    \subfloat[][$\mu$ = 0.5]{\includegraphics[width=0.9\columnwidth]{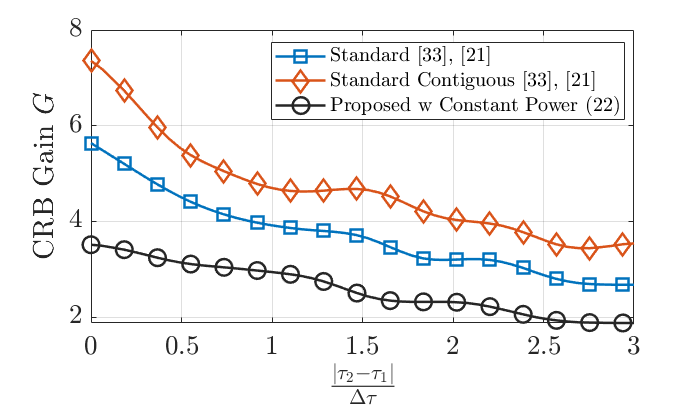}\label{subfig:CRBGain_050}} \\ \vspace{-0.2cm}
    \subfloat[][$\mu$ = 1]{\includegraphics[width=0.9\columnwidth]{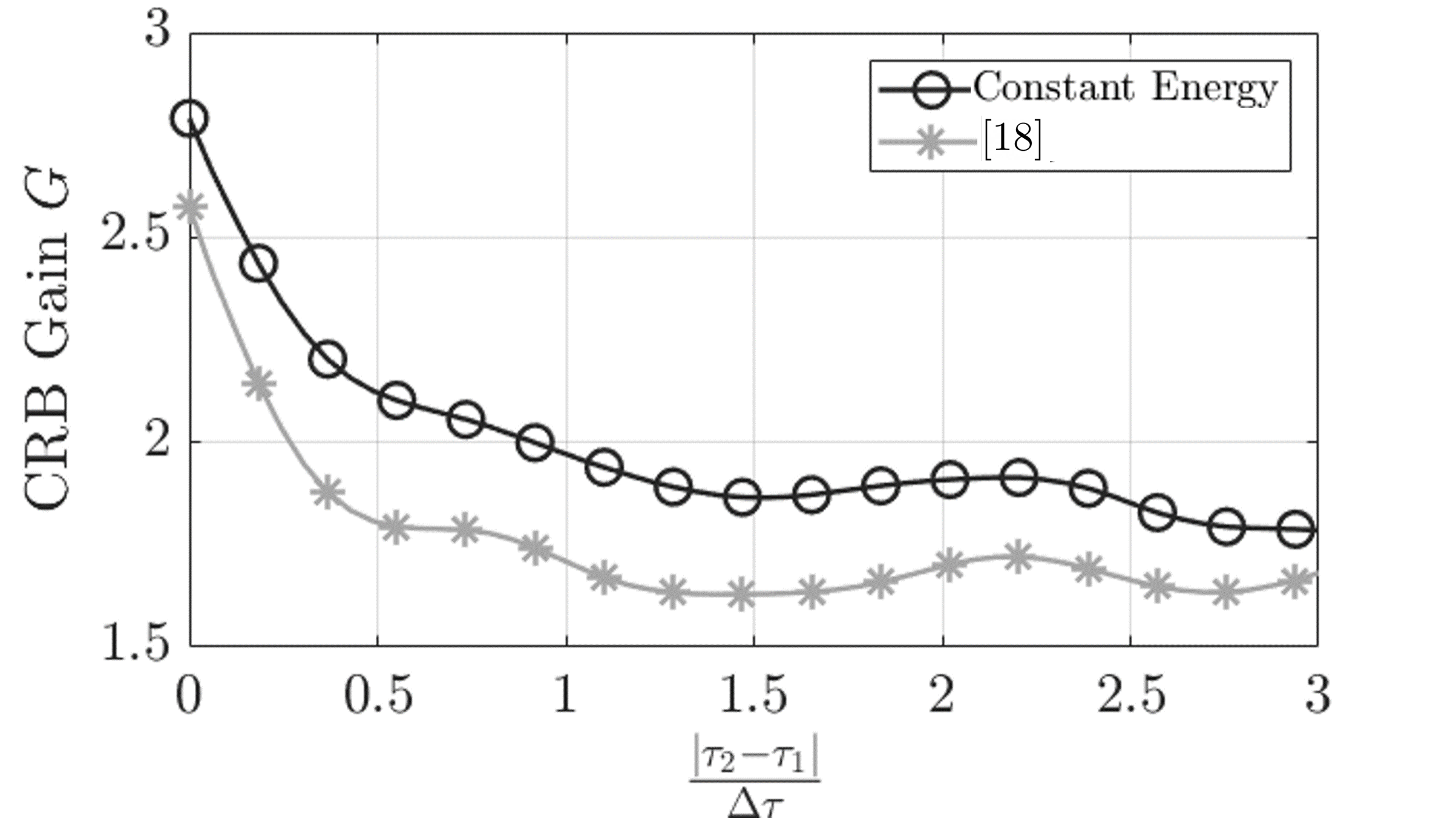}\label{subfig:CRBGain_1}}
    \caption{Sensing performance comparison between benchmarks in terms of CRB gain on delay estimation by changing the ROF $\mu$ }
    \label{fig:Performance_CRB}
\end{figure}
\section{Numerical Results}\label{sect:numerical_results}

This section evaluates the proposed ISAC waveform performance compared to two benchmarks: the standard-compliant random resource scheduling and the random scheduling with contiguous resources\cite{9345798}. Both benchmarks operate under a fixed occupancy factor ($\mu$) and maintain a constant energy per resource across time and frequency. The main difference lies in their allocation methods: the standard-compliant random scheduling assigns individual resources randomly, while the random contiguous scheduling assigns blocks of $N_b$ contiguous resources. Both benchmarks utilize linear interpolation for filling empty resources in time and frequency domains, as outlined in \cite{BarnetoFullDuplex}. The two aforementioned benchmarks pertain to bandwidths that are not fully occupied, where $\mu < 1$. To evaluate the proposed waveform under full bandwidth occupancy ($\mu = 1$), we compare it with a waveform from \cite{Wymeersch2021}, tailored for scenarios with two closely positioned targets.
A comparison is made between the two proposed waveforms: one with constant energy across time and frequency \eqref{eq:optProb1}, and the other that jointly optimizes energy-time-frequency (see \eqref{eq:optProb2}). This analysis highlights the benefits of integrating energy considerations into waveform design. The simulation parameters, unless otherwise indicated, are shown in Table \ref{tab:SimParam}.

The first result encompasses the CRB gain for estimating the delay between two closely located targets, defined as
\begin{align}
G = \frac{\mathrm{tr}(\mathbf{C}_{\tau, \text{b}})}{\mathrm{tr}(\mathbf{C}_{\tau, \text{opt}})} ,
\end{align}
where $\mathbf{C}_{\tau, \text{b}}$ represents the CRB matrix obtained via state-of-the-art methodologies in \cite{9345798} and \cite{Wymeersch2021}, and \eqref{eq:optProb1}, while $\mathbf{C}_{\tau, \text{opt}}$ is the CRB matrix obtained by \eqref{eq:optProb2}. A gain $G>1$ signifies a decrease in the CRB when employing the proposed time-frequency-energy optimized waveform compared to the benchmarks, indicating practical utility.  Doppler estimation CRB yields analogous outcomes and is omitted here for simplicity. Figure \ref{fig:Performance_CRB} shows the trend of the CRB gain $G$ in linear scale by varying the inter-delay spacing between the two targets, defined as $|\tau_2 - \tau_1|/\Delta \tau$ (quasi normalized to the delay resolution $\Delta \tau$) for different bandwidth occupancy $\mu = 0.25, 0.5$ and $1$ by considering $\Delta = 0$ dB. With the bandwidth occupancy factor $\mu = 0.25$ in Fig \ref{subfig:CRBGain_025}, the proposed ISAC waveform outperforms the standard-compliant random and random contiguous scheduling by obtaining a CRB gain $6 \times$ and $14\times$, respectively, when the two targets are closely placed. This underscores the effectiveness of the proposed waveform in discerning closely spaced targets. 
Moreover, a notable enhancement compared to both standard-compliant random and random contiguous scheduling, approximately $7 \times$ and $5\times$ respectively, is observed at $\mu = 0.5$ in Fig. \ref{subfig:CRBGain_050}. In contrast, Fig. \ref{subfig:CRBGain_1} denotes that the approach in \cite{Wymeersch2021} experiences limitations in minimizing the CRB due to the constraint in the ambiguity sidelobe levels, which forces the time-frequency resources to be more evenly spread within the bandwidth. Ultimately, the comparison between the energy-frequency-time optimized waveform (obtained by \eqref{eq:optProb2}) and the waveform optimized with constant energy  (obtained by \eqref{eq:optProb1}) is conducted. The average enhancements of a factor of 4 ($\mu= 0.25$ in Fig.\ref{subfig:CRBGain_025}), 3 ($\mu= 0.5$ in Fig.\ref{subfig:CRBGain_050}), and 2 ($\mu= 1$ in Fig.\ref{subfig:CRBGain_1}) respectively underscore the clear advantage of integrating energy optimization in defining the waveform. The oscillations visible in the graph are due to the ambiguity sidelobes.
\begin{figure}[b!]
    \centering
    \subfloat[][$\mu$ = 0.25]{\includegraphics[width=0.9\columnwidth]{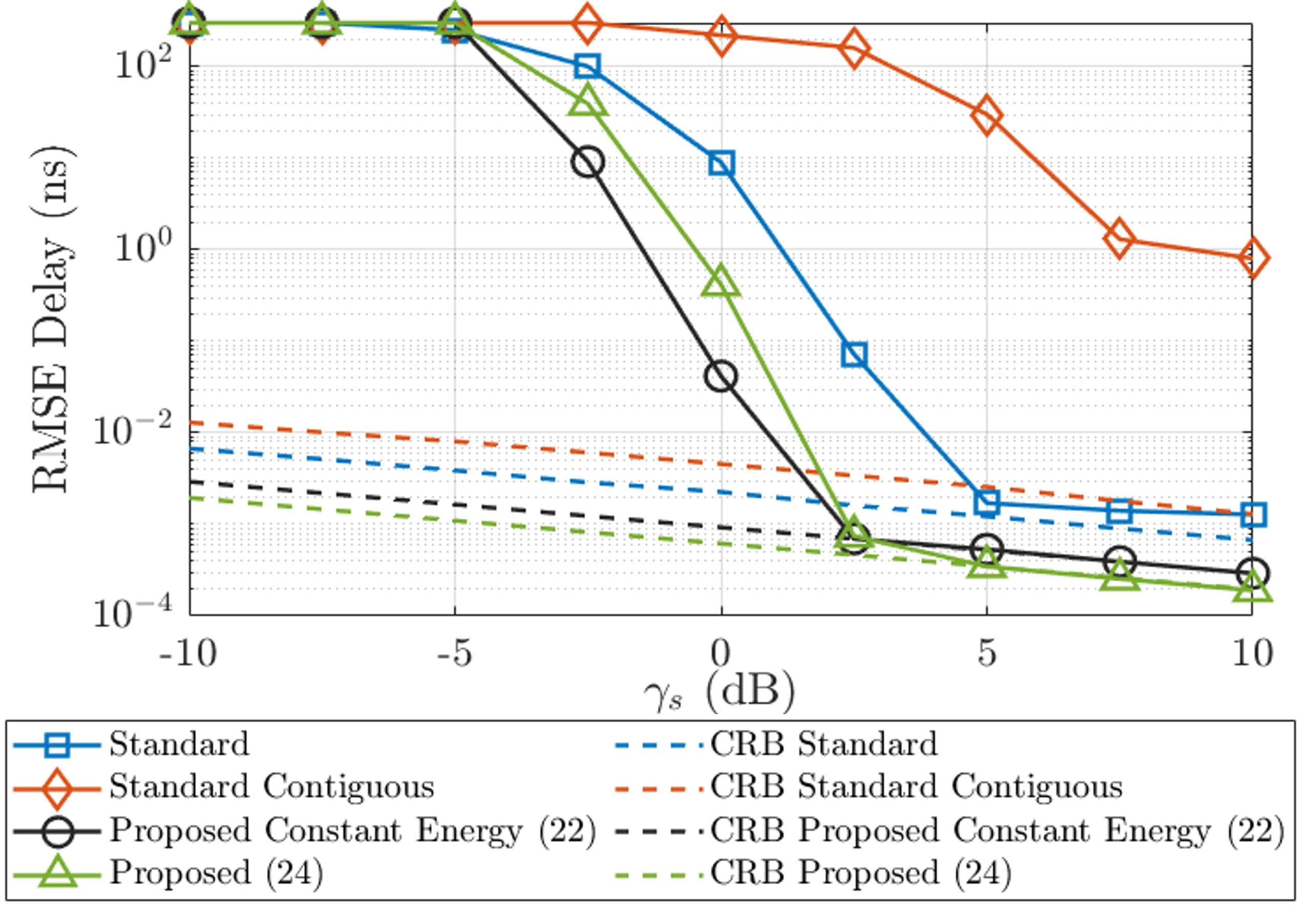}\label{subfig:ErrGain_mu05}}\\ \vspace{-0.2cm}
    \subfloat[][$\mu$ = 0.5]{\includegraphics[width=0.9\columnwidth]{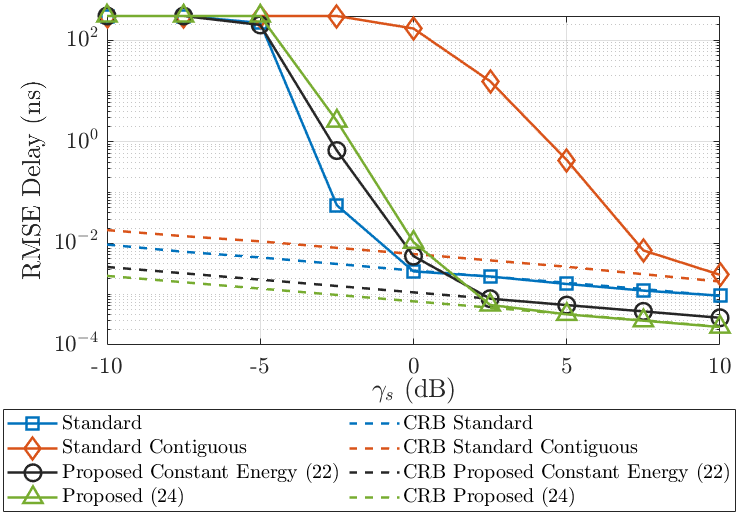}\label{subfig:ErrGain_mu1}}\\ \vspace{-0.2cm} 
    \subfloat[][$\mu$ = 1]{\includegraphics[width=0.9\columnwidth]{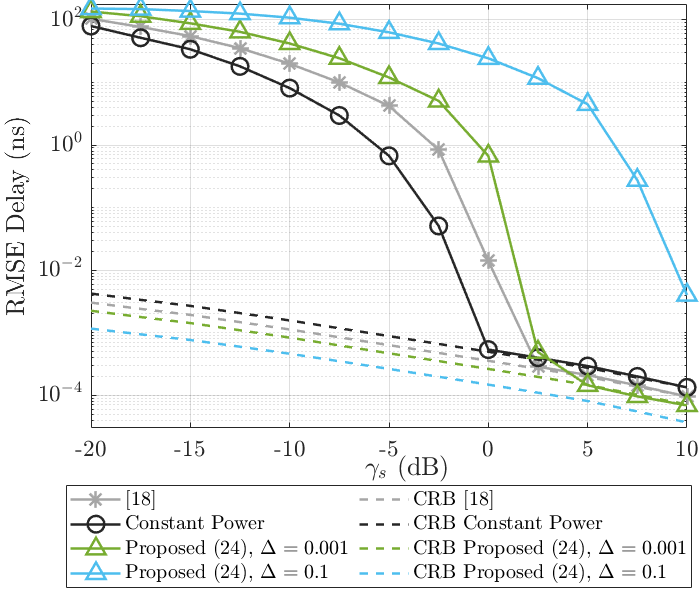}\label{subfig:ErrGain_mu1}}\\ \vspace{-0.2cm}
    \caption{Sensing performance comparison between benchmarks in terms of delay RMSE by changing the ROF $\mu$ }
    \label{fig:Performance_RMSE}
\end{figure}
Figure \ref{fig:Performance_RMSE} illustrates the root mean square error (RMSE) for the delay estimation achieved through the proposed methods and benchmark approaches, considering $\mu = 0.25$, $\mu = 0.5$ and $\mu = 1$ for total energy budget $E_{\text{tot}}$ = 43T dBmJ. The results of the proposed waveform in \eqref{eq:optProb2} are obtained with $\Delta = -15$ dB in the case of $\mu=0.25,0.5$ and $\Delta = -15,0$ dB if $\mu =1$, respectively.

The proposed waveform (both \eqref{eq:optProb1} and  
 \eqref{eq:optProb2}) notably enhances the CRB for delay estimation. As the average sensing SNR in the DD domain, defined as
\begin{align}
 \gamma_s = \frac{[\mathbf{e}_k \odot \mathbf{a}_k]_\ell \, |H_s|^2 MN}{N_0}
\end{align}
with $ |H_{s}|^2 = \mathbb{E}_\beta \left[\| \mathbf{H}_s \|_F^2\right]/L$, increases, the delay error approaches the CRB. At $\mu = 0.25$, the proposed interpolation method facilitates reaching the CRB at higher ${\gamma}_{s}$, thereby enhancing the performance as compared to standard random and random contiguous resource allocation methods employing the linear interpolation \cite{BarnetoFullDuplex}. The random contiguous scheduling struggles to attain the CRBs, indicating a failure of linear interpolation in this context. Conversely, the random waveform, allocating resources individually, demonstrates satisfactory performance solely at $\mu = 0.5$, ensuring adequate samples for linear interpolation. The numerical outcomes underscore the estimation enhancements provided by the proposed waveforms (both \eqref{eq:optProb1} and 
\eqref{eq:optProb2}) under stringent bandwidth occupancy constraints ($\mu = 0.25$), whereas for $\mu = 0.5$, standard random resource allocation with linear interpolation ensures commendable performance at low-medium levels of the SNR. However, for high sensing SNR levels, the proposed waveforms with constant and optimized energy improve the benchmarks.

Similar observations apply to Figure \ref{subfig:ErrGain_mu1}, where the proposed time-frequency-energy optimized waveform in \eqref{eq:optProb2} is compared with both the constant energy waveform and the approach outlined in \cite{Wymeersch2021}, opportunely adapted for the two targets scenario. 
The constant energy waveform demonstrates superior performance concerning RMSE, surpassing the method outlined in \cite{Wymeersch2021}. The proposed optimized waveform surpasses the performance of \cite{Wymeersch2021}, particularly in scenarios with high sensing SNR and a low value of $\Delta$. Consequently, excessively high values of $\Delta$ can enhance the CRB but simultaneously lead to elevated sidelobe levels. These high sidelobe levels have the potential to obscure weaker targets, thereby diminishing the estimation performance. Similar outcomes are observed for Doppler estimation, but it is omitted here for brevity.

\begin{figure}[t!]
    \centering
    \includegraphics[width=0.9\columnwidth]{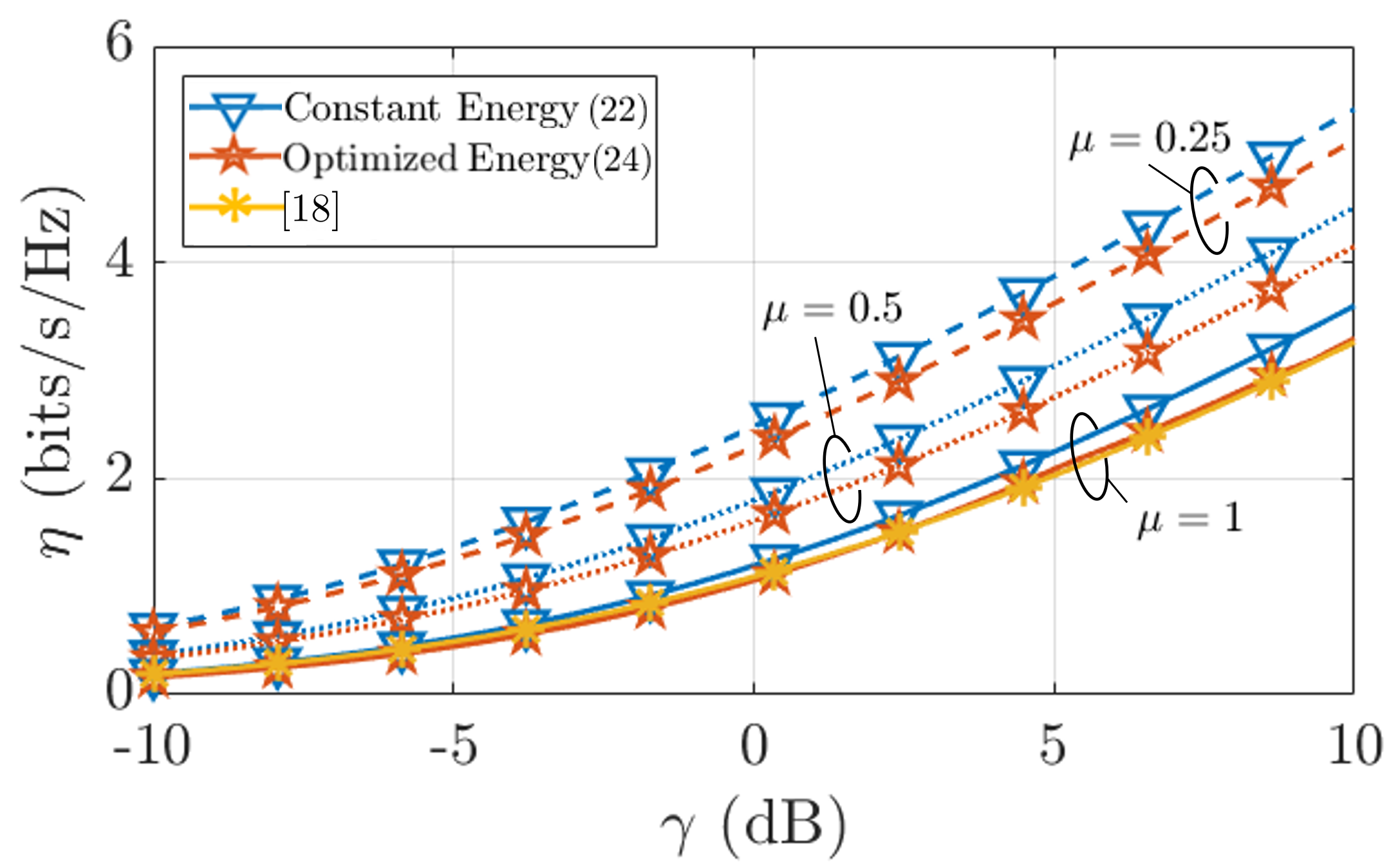}
     \caption{Achievable average rate obtained by the considered benchmarks and the proposed waveform  for different ROF $\mu$}\label{fig:se}
\end{figure}

The attainable SE is evaluated as depicted in Fig. \ref{fig:se} for both the optimized waveform with $\Delta = -15$ dB and benchmark cases, with $E_{\mathrm{tot}} = \mathbf{e}^{T}\mathbf{1}_{NM} = 43\times T$ dBmJ by varying the average communication SNR $\gamma$ defined as $\mathbb{E}_{k,\ell} [\,\gamma_{k,\ell}\,]$. The analysis reveals that enhancing the energy allocation at the extremities of the bandwidth results in comparable performance to that of constant energy waveforms when considering $\mu =0.25, 0.5$, as well as \cite{Wymeersch2021} for $\mu = 1$ when considering $\Delta = -15$ dB. The reduction in average SE resulting from an increase in the ROF $\mu$ occurs because, while maintaining the total energy constant, the SNR per resource bin diminishes with the ROF $\mu$ increase.
\begin{figure}[t!] 
    \centering
    \includegraphics[width=0.9\columnwidth]{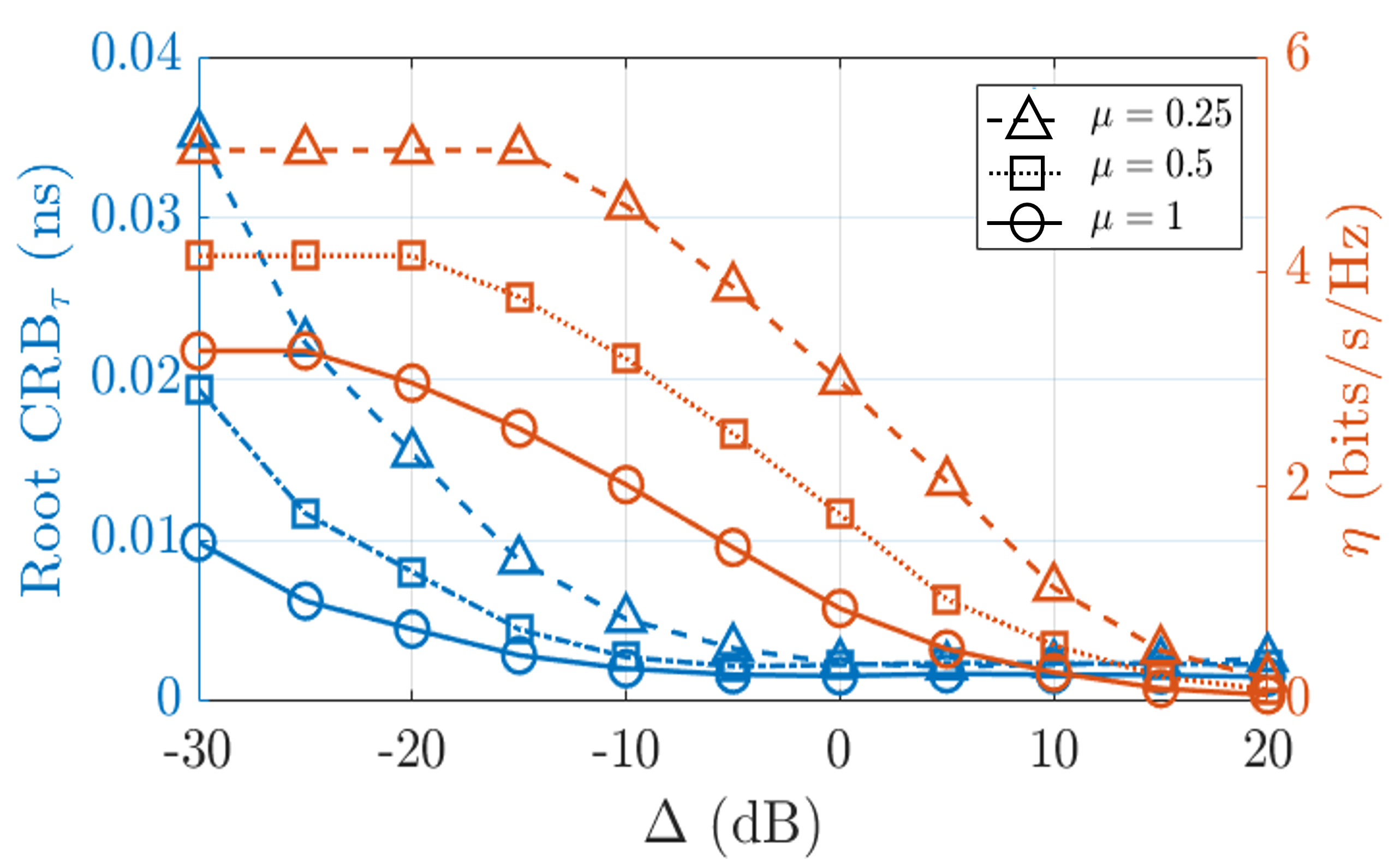}
     \caption{Trade-off between achievable average rate and root delay CRB in dependence of the energy gradient $\Delta$, obtained by the proposed waveform for different ROF $\mu$}\label{fig:trade_off}
\end{figure}

The SE is further analyzed in Fig. \ref{fig:trade_off}  by keeping constant the communication SNR $\gamma$, equal to 0 dB, while varying the energy gradient $\Delta$, which directly influences the energy distribution within the proposed waveform. In particular, Fig. \ref{fig:trade_off} shows the trade-off between the CRB and SE through variation in the gradient parameter $\Delta$. 
By fixing the total energy $E_{\text{tot}}$ a notable trend is visible: an increase in the energy gradient yields a reduction in the delay CRB, consequently enhancing sensing capabilities. Conversely, this increase in the energy gradient induces a proportional decrease in SE, thereby compromising communication performance. This observation underscores the importance of fine-tuning the energy gradient parameter to achieve satisfactory performance for both sensing and communication purposes.

\section{Conclusion}\label{sect:conclusion}

This paper introduces a novel ISAC waveform designed to optimize both communication and sensing performance via delay and Doppler weighted CRB minimization CRB, with achievable rate constraints. Unlike state-of-the-art methods, this design incorporates the ROF to more accurately reflect realistic time-frequency bandwidth usage. In practices, the resources are not fully occupied and lead to significant sidelobes in the ISAC waveform's ambiguity function, which impact target detection. To address this issue, we employ an interpolation technique based on the Schatten $p$-quasi norm, which requires fewer samples than the traditional nuclear norm, leading to more effective sensing channel interpolation in case of low ROFs. Numerical results demonstrate the superiority of the proposed ISAC method over standard random resource scheduling, showing a CRB gain of 6× and 14× compared to random and random contiguous scheduling, respectively, under ROF constraint. Furthermore, our approach achieves the CRBs in high SNR, while conventional methods fail. As a further contribution, we also detail and discuss the effect of enforcing a smooth energy variation within the resource grid, leading to an improved data rate.   

\appendices

\section{CRB on delay and Doppler for $K = 2$ targets}\label{app:CRB}
The CRB evaluation for the two targets case can be accomplished by using the matrix inversion lemma as
\begin{equation}
\mathbf{C}_{\tau} = \mathbf{J}_\tau^{-1},\,\, \mathbf{C}_{\nu} = \mathbf{J}_\nu^{-1}
\end{equation}
where $\mathbf{J}_\tau = \mathbf{F}_{\tau}-\mathbf{F}_{\nu,\tau}^{\mathrm{T}}\mathbf{F}_{\nu}^{-1} \mathbf{F}_{\nu,\tau}$ and $\mathbf{J}_\nu = \mathbf{F}_{\nu}-\mathbf{F}_{\nu,\tau}^{\mathrm{T}}\mathbf{F}_{\tau}^{-1} \mathbf{F}_{\nu,\tau}$ represent the delay and Doppler Schur complements, respectively. By considering two closely spaced targets of similar reflectivity, i.e., $|\beta_1|^2  \approx |\beta_2|^2$, the entries of the delay Schur complements are obtained as \eqref{eq:schur1}, \eqref{eq:schur2}, while the Doppler ones are obtained by replacing $\boldsymbol{\zeta}_\tau^{(i,j)}$ with $\boldsymbol{\zeta}_\nu^{(i,j)}$ and vice versa, with $i,j \in \{1,2\}$. Moreover, $[\mathbf{J}_{\tau}]_{1,1} \approx [\mathbf{J}_{\tau}]_{2,2}$, $[\mathbf{J}_{\nu}]_{1,1} \approx [\mathbf{J}_{\nu}]_{2,2}$.
\begin{figure*}
\begin{equation}\label{eq:schur1}
   [\mathbf{J}_{\tau}]_{1,1} =  \mathbf{e}^T \boldsymbol{\zeta}^{(1,1)}_{\tau}-\frac{\,|\mathbf{e}^T \boldsymbol{\zeta}^{(1,1)}_{\tau,\nu}|^2 \mathbf{e}^T \boldsymbol{\zeta}^{(1,1)}_{\nu}-2(\mathbf{e}^T \boldsymbol{\zeta}^{(1,2)}_{\tau,\nu} )(\mathbf{e}^T \boldsymbol{\zeta}^{(1,2)}_{\nu})(\mathbf{e}^T \boldsymbol{\zeta}^{(1,1)}_{\tau,\nu})+|\mathbf{e}^T \boldsymbol{\zeta}^{(1,2)}_{\tau,\nu}|^2\mathbf{e}^T \boldsymbol{\zeta}^{(1,1)}_{\nu}}{|\mathbf{e}^T \boldsymbol{\zeta}^{(1,1)}_{\nu}|^2-|\mathbf{e}^T \boldsymbol{\zeta}^{(1,2)}_{\nu}|^2},
\end{equation}
\begin{equation}\label{eq:schur2}
  [\mathbf{J}_{\tau}]_{1,2}= \mathbf{e}^T \boldsymbol{\zeta}^{(1,2)}_{\tau}-\frac{2(\mathbf{e}^T \boldsymbol{\zeta}^{(1,1)}_{\tau,\nu})(\mathbf{e}^T \boldsymbol{\zeta}^{(1,2)}_{\tau,\nu})(\mathbf{e}^T \boldsymbol{\zeta}^{(1,1)}_{\nu})-|\mathbf{e}^T \boldsymbol{\zeta}^{(1,1)}_{\tau,\nu}|^2(\mathbf{e}^T \boldsymbol{\zeta}^{(1,2)}_{\nu})-|\mathbf{e}^T \boldsymbol{\zeta}^{(1,2)}_{\tau,\nu}|^2(\mathbf{e}^T \boldsymbol{\zeta}^{(1,2)}_{\nu})}{|\mathbf{e}^T \boldsymbol{\zeta}^{(1,1)}_{\nu}|^2-|\mathbf{e}^T \boldsymbol{\zeta}^{(1,2)}_{\nu}|^2},
\end{equation}
\hrulefill
\end{figure*}
Therefore
\begin{equation}\label{eq: CRBtau1}
    [\mathbf{C}_{\tau}]_{1,1} = \frac{[\mathbf{J}_{\tau}]_{2,2}
    }{[\mathbf{J}_{\tau}]_{1,1}[\mathbf{J}_{\tau}]_{2,2}-([\mathbf{J}_{\tau}]_{1,2})^2}
\end{equation}
\begin{equation}\label{eq: CRBtau2}
    [\mathbf{C}_{\tau}]_{2,2} = \frac{[\mathbf{J}_{\tau}]_{1,1}
    }{[\mathbf{J}_{\tau}]_{1,1}[\mathbf{J}_{\tau}]_{2,2}-([\mathbf{J}_{\tau}]_{1,2})^2}
\end{equation}
The same procedure is applied to achieve $[\mathbf{C}_{\nu}]_{1,1}$ and $[\mathbf{C}_{\nu}]_{2,2}$.

\section{Relaxation of problem \eqref{eq:optProb1} and \eqref{eq:optProb2} }\label{app:OptimizationCRB}

The optimization problems specified in \eqref{eq:optProb1} and \eqref{eq:optProb2} contain a non-convex objective function, requiring manipulations to reformulate it into a conventional convex program.

\subsection{Objective Function: weighted sum of CRBs}
The weighted sum of the CRBs can be redefined in relation to the entries of the Schur complement in the following manner:
\begin{align}
  \epsilon_{\tau} \frac{ 2[\mathbf{J}_{\tau}]_{1,1}}{([\mathbf{J}_{\tau}]_{1,1})^2-([\mathbf{J}_{\tau}]_{1,2})^2} +  \epsilon_{\nu} \frac{2 [\mathbf{J}_{\nu}]_{1,1}}{([\mathbf{J}_{\nu}]_{1,1})^2-([\mathbf{J}_{\nu}]_{1,2})^2}.
\end{align}
As the denominator impacts more than the numerator term, the CRB can be represented utilizing the subsequent upper bound:
\begin{align}
  \epsilon_{\tau} \frac{2 x_{\tau}}{x_{\tau}^2-{j_{\tau}^2}} +  \epsilon_{\nu} \frac{2 x_{\nu}}{x_{\nu}^2-{j_{\nu}^2}},
\end{align}  
where $x_{\tau}, j_{\tau}$ are two auxiliary variables $\in \mathbb{R}^{+}$ such that
\begin{align}\label{eq:shur_1}
    x_{\tau} \leq [\mathbf{J}_{\tau}]_{1,1},\hspace{0.5cm}j_{\tau} \geq [\mathbf{J}_{\tau}]_{1,2}.
\end{align}
A new auxiliary variable $t_{\tau}$ is introduced such that the SOCP constraint is satisfied
\begin{equation}
   t_{\tau}x_{\tau} \geq {j_{\tau}^2}.
\end{equation}
Similar constraints are imposed on the variables $x_{\nu}$, $j_{\nu}$ and $t_{\nu}$ .
For both the delay and Doppler variables, the objective function is expressed as:
\begin{align}
    \frac{2 \epsilon_{\tau}}{x_{\tau}-t_{\tau}} + \frac{2 \epsilon_{\nu}}{x_{\nu}-t_{\nu}}.
\end{align}
Since the minimization of the initial objective function corresponds to maximizing its inverse, the objective function results in
\begin{align}
   \frac{\epsilon_{\nu} (x_{\tau} - t_{\tau}) + \epsilon_{\nu}(x_{\nu} - t_{\nu})}{4\epsilon_{\tau}\epsilon_{\nu}} -s,
\end{align}
where the variable $s \in \mathbb{R}^+$ is defined such that
\begin{align}
    s \geq \frac{\epsilon_{\nu}^2 (x_{\tau}-t_{\tau})^2 + \epsilon_{\tau}^2 (x_{\nu}-t_{\nu})^2}{4 \epsilon_{\tau}\epsilon_{\nu}(\epsilon_{\nu}(x_{\tau}- t_{\tau})+ \epsilon_{\tau}(x_{\nu}- t_{\nu}))}.
\end{align}

Hence, the optimization problem in \eqref{eq:optProb1} can be reformulated as a MICP model as
\begin{subequations}\label{eq:2targ_ref}
\begin{alignat}{2} 
&\underset{\substack{\mathbf{a}_k, \mathbf{e}_k,t_{\tau},t_{\nu},\\x_{\tau}, x_{\nu},j_{\tau}, j_{\nu}, s}}{\text{max}}  &\quad&    \frac{\epsilon_{\nu} (x_{\tau} - t_{\tau}) + \epsilon_{\nu}(x_{\nu} - t_{\nu})}{4\epsilon_{\tau}\epsilon_{\nu}} -s\label{eq:optk2Prob_r}\\
&\text{s.t.} &    &\eqref{eq:prob1_constraint1},\eqref{eq:prob1_constraint2},\eqref{eq:prob1_constraint3} \nonumber\\
&  &      & \left\lvert\left\lvert\begin{bmatrix} &\sqrt{2}j_{\tau}\\&x_{\tau}\\&t_{\tau}\end{bmatrix} \right\lvert\right\lvert_2 \leq x_{\tau} + t_{\tau}\label{eq:prob4_constraint2},\\
&  &      & \left\lvert\left\lvert\begin{bmatrix} &\sqrt{2}j_{\nu}\\&x_{\nu}\\&t_{\nu}\end{bmatrix}\right\lvert\right\lvert_2 \leq  x_{\nu} + t_{\nu}\label{eq:prob4_constraint2},\\
&  &      & \left\lvert\left\lvert\begin{bmatrix} &\epsilon_{\tau}(x_{\nu}-t_{\nu})\\&\epsilon_{\nu}(x_{\tau}-t_{\tau})\\&2\epsilon_{\tau}\epsilon_{\nu}s\\&\epsilon_{\nu}(x_{\tau}-t_{\tau}) +\epsilon_{\tau}(x_{\nu}-t_{\nu})\end{bmatrix} \right\lvert\right\lvert_2 \leq\nonumber\\ 
&   &     &2\epsilon_{\tau}\epsilon_{\nu} s +\epsilon_{\nu}(x_{\tau}-t_{\tau}) +\epsilon_{\tau}(x_{\nu}-t_{\nu})\label{eq:prob4_constraintt},\\
&  &      &x_{\tau} \geq t_{\tau},\,\,x_{\nu} \geq t_{\nu},\\
&  &      &x_{\tau} \leq [\mathbf{J}_{\tau}]_{1,1},\,\,x_{\nu} \leq [\mathbf{J}_{\nu}]_{1,1},\label{eq:shurC1}\\
&  &      &j_{\tau} \geq [\mathbf{J}_{\tau}]_{1,2},\,\,j_{\nu} \geq [\mathbf{J}_{\nu}]_{1,2}\label{eq:shurC2}.
\end{alignat}
\end{subequations}
The problem expressed in \eqref{eq:optProb2} can be reformulated analogously.
 
\subsection{Shur Complement Constraints}

The constraints delineated in \eqref{eq:shurC1} and \eqref{eq:shurC2} require mathematical computation. The former is rewritten as
\begin{align}\label{eq:constraint2target}
&(\mathbf{e}^\mathrm{T} \boldsymbol{\zeta}^{(1,1)}_{\tau}-x_{\tau})(\mathbf{e}^\mathrm{T} \boldsymbol{\zeta}^{(1,1)}_{\nu}+\mathbf{e}^\mathrm{T} \boldsymbol{\zeta}^{(1,2)}_{\nu}) \geq  \\&|\mathbf{e}^\mathrm{T} \boldsymbol{\zeta}^{(1,1)}_{\tau,\nu}|^2 + |\mathbf{e}^\mathrm{T} \boldsymbol{\zeta}^{(1,2)}_{\tau,\nu}|^2 +\nonumber\\& \frac{|\mathbf{e}^\mathrm{T} \boldsymbol{\zeta}^{(1,1)}_{\tau,\nu}-\mathbf{e}^\mathrm{T} \boldsymbol{\zeta}^{(1,2)}_{\tau,\nu} |^2\,\mathbf{e}^\mathrm{T} \boldsymbol{\zeta}^{(1,2)}_{\nu} }{\mathbf{e}^\mathrm{T} \boldsymbol{\zeta}^{(1,1)}_{\nu} -\mathbf{e}^\mathrm{T} \boldsymbol{\zeta}^{(1,2)}_{\nu}} \nonumber,
\end{align}
which, according to the definition of $\boldsymbol{\zeta}_{\nu}$, $\boldsymbol{\zeta}_{\tau}$ and $\boldsymbol{\zeta}_{\tau,\nu}$ can be relaxed to 
\begin{align}\label{eq:constraint2target}
&(\mathbf{e}^\mathrm{T} \boldsymbol{\zeta}^{(1,1)}_{\tau}-x_{\tau})(\mathbf{e}^\mathrm{T} \boldsymbol{\zeta}^{(1,1)}_{\nu}+\mathbf{e}^\mathrm{T} \boldsymbol{\zeta}^{(1,2)}_{\nu}) \geq \\&|\mathbf{e}^\mathrm{T} \boldsymbol{\zeta}^{(1,1)}_{\tau,\nu}|^2 + |\mathbf{e}^\mathrm{T} \boldsymbol{\zeta}^{(1,2)}_{\tau,\nu}|^2 +\nonumber\\& |\mathbf{e}^\mathrm{T} \boldsymbol{\zeta}^{(1,1)}_{\tau,\nu}-\mathbf{e}^\mathrm{T} \boldsymbol{\zeta}^{(1,2)}_{\tau,\nu} |^2 \nonumber,
\end{align}
and it can be easily recast as a more stringent SOCP constraint by ensuring $\mathbf{e}^\mathrm{T} \boldsymbol{\zeta}^{(1,1)}_{\tau}-x_{\tau} \geq 0$.

The same procedure can be applied to  the constraint in \eqref{eq:shurC2} as:
\begin{align}
\begin{split}
&(j_{\tau}-\mathbf{e}^\mathrm{T} \boldsymbol{\zeta}^{(1,2)}_{\tau})(\mathbf{e}^\mathrm{T} \boldsymbol{\zeta}^{(1,1)}_{\nu}-\mathbf{e}^\mathrm{T} \boldsymbol{\zeta}^{(1,2)}_{\nu})\geq \\&|\mathbf{e}^\mathrm{T} \boldsymbol{\zeta}^{(1,1)}_{\tau,\nu}|^2 + |\mathbf{e}^\mathrm{T} \boldsymbol{\zeta}^{(1,2)}_{\tau,\nu}|^2 -\nonumber\\& \frac{|\mathbf{e}^\mathrm{T} \boldsymbol{\zeta}^{(1,1)}_{\tau,\nu}+\mathbf{e}^\mathrm{T} \boldsymbol{\zeta}^{(1,2)}_{\tau,\nu} |^2\,\mathbf{e}^\mathrm{T} \boldsymbol{\zeta}^{(1,1)}_{\nu} }{\mathbf{e}^\mathrm{T} \boldsymbol{\zeta}^{(1,1)}_{\nu} +\mathbf{e}^\mathrm{T} \boldsymbol{\zeta}^{(1,2)}_{\nu}}.
\end{split}
\end{align}

which, by ensuring $j_{\tau}-\mathbf{e}^\mathrm{T} \boldsymbol{\zeta}^{(1,2)}_{\tau} \geq 0$, can be simplified in the following  more stringent SOCP constraint
\begin{align}
\begin{split}
&(j_{\tau}-\mathbf{e}^\mathrm{T} \boldsymbol{\zeta}^{(1,2)}_{\tau})(\mathbf{e}^\mathrm{T} \boldsymbol{\zeta}^{(1,1)}_{\nu}-\mathbf{e}^\mathrm{T} \boldsymbol{\zeta}^{(1,2)}_{\nu}) \geq \\&|\mathbf{e}^\mathrm{T} \boldsymbol{\zeta}^{(1,1)}_{\tau,\nu}|^2 + |\mathbf{e}^\mathrm{T} \boldsymbol{\zeta}^{(1,2)}_{\tau,\nu}|^2.
\end{split}
\end{align}
A comparable procedure is employed for the Schur complement constraints associated with the Doppler. 

\subsection{Spectral Efficiency Constraint}
The QoS rate constraint outlined in \eqref{eq:prob1_constraint1} exhibits convexity and can be redefined as an exponential cone constraint through the introduction of an auxiliary optimization variable $\mathbf{y}_k \in \mathbb{R}_{+}^{L \times 1},\forall k $ such that:
\begin{align}\label{eq:rate_c}
\begin{split}
&[\mathbf{y}_k]_\ell \leq \log_2(1+\gamma_{k,\ell}), \forall k, \forall \ell\\
&\xi\,\,[\mathbf{e}_k]_\ell +1 \geq \exp(\ln(2)\,[\mathbf{y}_k]_\ell), \forall k,\forall \ell
\end{split}
\end{align}
with $\xi = \frac{|H_{k}|^2}{N_0}$.  When considering only the time-frequency optimization, $\mathbf{e}_k = \sigma^2 \mathbf{a}_k$. The SE threshold is than achieved by the constraint
\begin{align}
    \frac{1}{L}\sum_{\ell=1}^{L}\,[\mathbf{y}_k]_{\ell} \geq  \overline{\eta},\,\,\forall k.
\end{align}

\bibliographystyle{IEEEtran}
\bibliography{Bibliography,Bibliography_TWC}

\end{document}